\documentclass[aps,prd,twocolumn,superscriptaddress]{revtex4-2}
\usepackage{amsmath}
\usepackage{amssymb}
\usepackage{epsfig}
\usepackage{cancel}
\usepackage{soul}
\usepackage{color, xcolor}
\usepackage{makecell}
\usepackage[colorlinks=true,linkcolor=red,urlcolor=blue,citecolor=blue]{hyperref}
\usepackage{verbatim}
\newcommand{\md}{\mathrm d}
\newcommand{\p}{\partial}
\begin{document}

\title{Inflation in light of ACT/SPT: A new perspective from Weyl gravity}

\author{Qing-Yang Wang}
\email{wangqy@ucas.ac.cn}
\affiliation{School of Fundamental Physics and Mathematical Sciences, Hangzhou Institute for Advanced Study, UCAS, Hangzhou 310024, China}
\date{\today}

\begin{abstract}
Recent measurements from the Atacama Cosmology Telescope (ACT) and the South Pole Telescope (SPT) have placed the strictest constraints on the primordial scalar perturbation spectrum, reporting a spectral index of $n_s\sim0.967-0.98$ at 95\% confidence level. This result indicates a stronger scale invariance of the scalar perturbation than earlier estimates, posing challenges for numerous inflation models. In this work, we propose a novel inflationary scenario based on Weyl scale-invariant gravity, where the quadratic curvature establishes the scale invariance of the scalar spectrum, while the higher-order extensions imprint the observed slight deviation. Specifically, the exponential curvature extensions are introduced to suppress the mass divergence of the inflaton. We find such scenario naturally yields leading-order predictions of $n_s\simeq1-3/(2N)\sim0.97-0.975$ or $n_s\simeq1-5/(3N)\sim0.967-0.972$ for various models, in excellent agreement with the ACT/SPT constraints. This result builds a concrete bridge between theoretical and observational scale invariance, implying an enduring cosmic echo of the primordial symmetry.

\end{abstract}

\maketitle

{\it Introduction} —
It is well established that the observations of cosmic microwave background (CMB) reveal an almost scale-invariant scalar perturbation spectrum, characterized by a spectral index $n_s$ slightly less than 1. Recently, the Atacama Cosmology Telescope (ACT) \cite{AtacamaCosmologyTelescope:2025blo,AtacamaCosmologyTelescope:2025nti} and the South Pole Telescope (SPT) \cite{SPT-3G:2025bzu} have released updated CMB measurements, indicating an $n_s$ value closer to 1 than previously reported in Planck 2018 \cite{Planck:2018vyg}, especially after incorporating the latest baryon acoustic oscillation data from DESI \cite{DESI:2025zgx}. Specifically, they constrain $n_s\sim0.967-0.98$ in the 95\% confidence interval. This suggests that the scale invariance of the scalar spectrum is even stronger than earlier estimates.

An important goal of modern cosmology is to understand the origin of this approximate scale invariance. Generally, inflation stands as a compelling theoretical framework capable of generating such a nearly scale-invariant spectrum \cite{Mukhanov:1981xt}. Over the years, numerous inflation models have been developed. However, with the increasing precision of CMB observations—particularly following the release of the ACT/SPT results—many of these models are now facing stringent tests. For instance, well-studied scenarios such as Starobinsky inflation \cite{Starobinsky:1980te}, Higgs inflation \cite{Bezrukov:2007ep}, and $\alpha$-attractor T models \cite{Kallosh:2013yoa} predict a value of $n_s$ that is now too low to be consistent with the latest CMB data, leaving them disfavored at the 95\% confidence level. These developments urge a reassessment of the inflationary mechanism, especially its intrinsic connection to scale invariance.

An enlightening idea posits that inflation could emerge from a gravity theory with local scale symmetry, also known as Weyl scale invariance. This symmetry was first introduced by H.~Weyl \cite{Weyl:1918ib,Weyl:1919fi}, and now has been widely applied in particle physics \cite{Cheng:1988zx,Scholz:2014kba,Ohanian:2015wva,Ferreira:2018itt,Ghilencea:2018dqd,Oda:2020cmi,Ghilencea:2021lpa}, cosmology \cite{Ferreira:2019zzx,Ghilencea:2019rqj,Ghilencea:2020piz,Tang:2020ovf,Cai:2021png,Smolin:1979uz,Nishino:2009in,Bars:2013yba,Quiros:2014hua,Ferreira:2016wem,Ferreira:2018qss,Tang:2018mhn,Ghilencea:2018thl,Tang:2019uex,Tang:2019olx,Ghilencea:2020rxc,Ghilencea:2021jjl,Wang:2022ojc,Wang:2023hsb,Hu:2023yjn,Burikham:2023bil,Bhagat:2023ych,Harko:2024fnt,Gialamas:2024iyu,Lee:2024rjw,Konoplya:2025mvj,Lalak:2025mil,Katsoulas:2025srh}, and quantum gravity \cite{Wu:2015wwa,Ghilencea:2024usf}. In such a framework of inflation, the spontaneous breaking of the Weyl symmetry would naturally give rise to a nearly scale-invariant scalar spectrum, implying a profound connection between the theoretical and observational scale invariance. 

With Weyl symmetry, if gravity is formulated in a purely geometric and ghost-free manner, the most general allowed action reduces to the quadratic curvature term $\hat R^2$, where $\hat R$ denotes the Ricci scalar in Weyl geometry. This model is equivalent to a cosmological constant in the Einstein frame \cite{Ghilencea:2018dqd}, corresponding to a pure de Sitter spacetime and a completely scale-invariant scalar spectrum, while, once matter is introduced, such as a scalar field, more scale-invariant terms are allowed in the action. They modify the induced cosmological constant to an effective scalar potential $V(\Phi)$ with asymptotically flat region in certain cases. This scenario is capable of realizing a viable inflation model and bringing about a slight deviation from the exact scale invariance of the spectrum present in the purely geometric case.

In previous studies, a scale-invariant linear curvature term $\phi^2\hat R$ is often introduced alongside the quadratic term to realize the above scenario \cite{Ferreira:2019zzx,Ghilencea:2019rqj,Ghilencea:2020piz,Tang:2020ovf,Cai:2021png}. This model leads to a Higgs-like potential for inflation. However, its predicted value for $n_s$ is lower than that of the Starobinsky model, thus further exacerbating the tension with the current observational constraint.

In this work, we explore the role of higher-order curvature extensions—rather than a linear term—in departing the pure de Sitter state derived from the $\hat R^2$ model. Such extensions are not simple corrections to prior models, but rather lead to a brand new inflationary scenario. Specifically, we consider exponential curvature extensions, which, by contrast with simple polynomial terms, significantly alleviate the mass divergence of inflaton. We calculate cosmological observables for such models and confirm their consistency with the ACT/SPT results.

These conventions are adopted in this work: Friedmann-Lemaître-Robertson-Walker metric $\md s^2=-\md t^2+a^2(t)\md\mathbf x^2$, natural unit $\hbar=c=1$, and Planck mass $M_P\equiv1/\sqrt{8\pi G}=2.435\times10^{18}\mathrm{GeV}$.

{\it Weyl gravity} —
We start with a brief introduction to the Weyl $F(\hat R,\phi)$ gravity, which contains a scalar $\phi$ and a Weyl gauge field $W_\mu\equiv g_Ww_\mu$ with $g_W$ the gauge coupling. The Lagrangian in the Jordan frame is given by
\begin{align}\label{LWeyl}
    \frac{\mathcal L_J}{\sqrt{-g}}=\frac{1}{2}F(\hat R,\phi)-\frac{1}{4g_W^2}F_{\mu\nu}F^{\mu\nu}-\frac{\zeta}{2}D_\mu\phi D^\mu\phi,
\end{align}
where $\zeta$ is a free parameter, $D_\mu=\p_\mu-W_\mu$ is the covariant derivative for scalar, $F_{\mu\nu}\equiv\p_\mu W_\nu-\p_\nu W_\mu$ is the field strength tensor of $W_\mu$, $\hat R$ denotes the Weyl Ricci scalar derived from the Weyl scale-invariant connection
\begin{align}\label{connection}
    \hat{\Gamma}^\rho_{\mu\nu}=\Gamma^\rho_{\mu\nu}+\left(W_\mu g^\rho_\nu+W_\nu g^\rho_\mu-W^\rho g_{\mu\nu}\right),
\end{align}
and $F(\hat R,\phi)$ is a combination of $\hat R$ and $\phi$ preserving the scale invariance of the Lagrangian. We introduce the following Weyl transformations,
\begin{align}
		&g_{\mu\nu}\to g'_{\mu\nu}=f^2(x)g_{\mu\nu},~\phi\to \phi'=f^{-1}(x)\phi,\nonumber\\
        &W_\mu\to W'_\mu=W_\mu-\p_\mu\ln|f(x)|.
\end{align}
One can derive $\hat R\to f^{-2}(x)\hat R$ and $\sqrt{-g}\to f^4(x)\sqrt{-g}$, then verify the invariance of Eqs.~(\ref{LWeyl}) and (\ref{connection}) under these transformations.

To obtain the Lagrangian in the Einstein frame, we introduce a dimension-2 auxiliary field $\chi$ obeying $\chi=\hat R$, and rewrite the curvature term as
\begin{align}
    F(\hat R,\phi)=F_{,\chi}(\chi,\phi)\left(\hat R-\chi\right)+F(\chi,\phi), 
\end{align}
where $F_{,\chi}(\chi,\phi)\equiv\frac{\p F(\hat R\to\chi,\phi)}{\p\chi}$. By setting up a specific Weyl transformation with $f(x)=\sqrt{F_{,\chi}(\chi,\phi)}/M_P$, which amounts to the gauge fixing condition
\begin{align}\label{fixgauge}
    F_{,\chi}(\chi,\phi)=M_P^2,
\end{align}
the Lagrangian transfers to the Einstein frame,
\begin{gather}
    \frac{\mathcal L_E}{\sqrt{-g}}=\frac{M_P^2}{2}\hat{R}-\frac{1}{4g_W^2}F_{\mu\nu}F^{\mu\nu}-\frac{\zeta}{2}D_{\mu}\phi D^{\mu}\phi-V(\phi),\\
    V(\phi)=-\frac{1}{2}\left[F(\chi,\phi)-\chi M_P^2\right].\label{Vphi}
\end{gather}
Clearly, the scale symmetry breaks spontaneously and an effective scalar potential emerges, where $\chi$ is expressed as a function of $\phi$ by solving Eq.~(\ref{fixgauge}).

Now we translate the Lagrangian to a more familiar form. Compared to the conventional Ricci scalar $R$, one can verify that
\begin{align}\label{WeylR}
	\hat{R}=R-6 W_{\mu} W^{\mu}-\frac{6}{\sqrt{-g}} \p_{\mu}\left(\sqrt{-g} W^{\mu}\right).
\end{align}
The total derivative term here is generally removed from the Lagrangian due to its null surface integral, while the second term can be combined with the $\phi$ kinetic term and together rewritten as
\begin{align}\label{phiandW}
    \frac{\zeta}{2}D_\mu&\phi D^\mu\phi+3M_P^2W_{\mu}W^{\mu}=\nonumber\\
    &\frac{1}{2}\frac{6M_P^2\zeta}{6M_P^2+\zeta\phi^2}\p_\mu\phi\p^\mu\phi+\frac{1}{2}\left(6M_P^2+\zeta \phi^2\right)\overline W^2,
\end{align}
where we have redefined the Weyl gauge field as
\begin{align}
    \overline W_\mu\equiv W_\mu-\frac{1}{2}\p_\mu\ln\left|6M_P^2+\zeta\phi^2\right|\equiv g_W\overline w_\mu.
\end{align}
Further redefining the scalar field
\begin{align}\label{phi2}
    \frac{\phi^2}{M_P^2} \equiv
    \begin{cases}
    \frac{6}{|\zeta|} \sinh ^{2}\left(\frac{ \pm \Phi}{\sqrt{6}M_P}\right),~\zeta>0, \\
    \frac{6}{|\zeta|} \cosh ^{2}\left(\frac{ \pm \Phi}{\sqrt{6}M_P}\right),~\zeta<0,
    \end{cases}
\end{align}
the mass of the Weyl gauge field is expressed as
\begin{align}\label{mw}
    m_W^2(\Phi)&=g_W^2\left(6M_P^2+\zeta\phi^2\right)\nonumber\\
    &=
    \begin{cases}
    +6g_W^2M_P^2\cosh^2\left(\frac{ \pm \Phi}{\sqrt{6}M_P}\right),~\zeta>0,\\
    -6g_W^2M_P^2\sinh^2\left(\frac{ \pm \Phi}{\sqrt{6}M_P}\right),~\zeta<0,
    \end{cases}
\end{align}
and the Lagrangian in the Einstein frame finally reduces to a conventional form,
\begin{align}\label{LEWeyl}
    \frac{\mathcal L_E}{\sqrt{-g}}=&\frac{M_P^2}{2}R-\frac{1}{2}\p_\mu\Phi \p^\mu\Phi-V(\Phi)\nonumber\\
    &-\frac{1}{4g_W^2}\overline F_{\mu\nu}\overline F^{\mu\nu}-\frac{1}{2}m_W^2(\Phi)\overline w_\mu\overline w^\mu.
\end{align}
As can be seen, it provides an inflaton candidate $\Phi$ and a vector dark matter candidate $\overline w_\mu$ coupled with $\Phi$. Since negative $\zeta$ leads to $m_W^2<0$, we only focus on the $\zeta>0$ case in this work.

{\it Models} —
In this framework, $F(\hat R,\phi)$ determines the configuration of the effective potential $V(\Phi)$, which can be derived from Eqs.~(\ref{fixgauge}), (\ref{Vphi}), and (\ref{phi2}), albeit the analytic expression may be unavailable for some complex cases. With these equations, one can readily verify the conclusions we mentioned earlier: the purely geometric case $F=\alpha\hat R^2$ is equivalent to a cosmological constant, and an additional scalar $\phi$ is able to make the model dynamical, leading to an inflation model.

A straightforward attempt is to introduce a polynomial extension,
\begin{align}\label{Fpoly}
    F(\hat R,\phi)=\alpha\hat R^2+\lambda_n\hat R^n\phi^{4-2n}.
\end{align}
As noted earlier, the $n=1$ case, which includes the linear curvature term, is widely investigated in previous studies. This model reduces to the purely geometric case (flat potential) for $\phi\to0$, where the scale-invariant spectrum forms. Inflation occurs during the evolution of $\phi$ from 0 to the potential minimum, which is $\phi=\pm1$ in this model. The deviation of $\phi$ from 0 when horizon crossing determines the deviation of the spectrum from scale invariance. Specifically, the effective potential is a Higgs-like potential,
\begin{align}\label{VWR2}
	V(\Phi)=
	\frac{M_P^4}{8\alpha}\left[1-\frac{6}{|\zeta|}\sinh^2\left(\frac{\Phi}{\sqrt{6}M_P}\right)\right]^2,~\zeta>0.
\end{align}
It predicts a spectral index $n_s\simeq n_s^*-11N/\zeta^2$ \cite{Wang:2023hsb}, where $n_s^*$ denotes the prediction in the Starobinksy model. This result, however, is now disfavored by ACT/SPT constraints at the 95\% confidence level. Therefore, we have to consider other possibilities.

Actually, the linear curvature term in the Jordan frame is unnecessary, as any form of $F(\hat R,\phi)$ can induce the linear curvature in the Einstein frame after gauge fixing. With this in mind, we ponder on certain higher-order extensions rather than the linear term in this work. One may try the $n\geq3$ case in Eq.~(\ref{Fpoly}). It reduces to the purely geometric case when $\phi\to\infty$ and creates a potential minimum at $\phi=0$. In other words, it develops a "pit" on the cosmological constant potential around $\phi=0$. Inflation occurs during the evolution of $\phi$ from a large value to 0, and meanwhile induces the deviation of the spectrum from scale invariance.

This model, however, is plagued by other problems in the vicinity of $\phi=0$. For the $n=3$ case, the asymptotic form of the effective potential, $V_{\Phi\to0}\propto|\Phi|M_P^3-\Phi^2M_P^2$, indicates that the inflaton's effective mass remains $m_\Phi^2=V''(\Phi)<0$ after inflation, resulting in a permanent tachyonic instability. For cases with $n\geq4$, the potential takes the form $V_{\Phi\to0}\propto|\Phi|^{2-\frac{2}{n-1}}$, which leads to a divergent effective mass, $m_\Phi^2\propto|\Phi|^{-\frac{2}{n-1}}$, as $|\Phi|$ decreases. Thus, when the inflaton field falls below a critical value $|\Phi_c|$, the effective mass surpasses the Planck scale, thereby rendering the effective description invalid.

To alleviate this problem, we propose that the $\hat R^2$ model with exponential extensions has better behavior in the vicinity of $\phi=0$ than the simple polynomial extensions. There are several example models:
\begin{align}
    &F(\hat R,\phi)=\alpha\hat R^2e^{\beta\hat R/\phi^2},\label{modele}\\
    &F(\hat R,\phi)=\alpha\phi^2\hat R\left(e^{\beta\hat R/\phi^2}-1\right),\label{modele-1}\\
    &F(\hat R,\phi)=\alpha\hat R^2\cosh\left(\beta\hat R/\phi^2\right),\label{modelch}\\
    &F(\hat R,\phi)=\alpha\phi^2\hat R\sinh\left(\beta\hat R/\phi^2\right),\label{modelsh}
\end{align}
where $\alpha$ and $\beta$ are positive parameters. Since $\beta$ couples exclusively to $\phi^2$ in the exponential term, it can be absorbed by $\zeta$ in Eq.~(\ref{phi2}) via parameter redefinition. Therefore, we will omit it in the following discussions. Through series expansion, one can confirm that these models reduce to the model (\ref{Fpoly}) at the sub-leading-order. Specifically, models (\ref{modele}) and (\ref{modele-1}) correspond to the $n=3$ case. We name this type of model the type A model. Models (\ref{modelch}) and (\ref{modelsh}) correspond to the $n=4$ case, which are classed as the type B model.

Note that the simplification of exponential-extended models with Eq.~(\ref{Fpoly}) is only valid in the regime $\phi^2\gg \hat R$. The behavior near $\phi=0$, particularly the divergence behavior, is presented in different manners. Since Eq.~(\ref{fixgauge}) in these models lacks explicit solutions, we construct an asymptotic solution in the limit $\phi\to0$. To leading order, all models yield the same estimate: $\chi_{\phi\to0}\simeq 2\phi^2\ln\left|\frac{M_P}{\phi}\right|$. Substituting this into Eq.~(\ref{Vphi}) gives
\begin{align}
    &V_{\phi\to0}\simeq\phi^2M_P^2\ln\left|\frac{M_P}{\phi}\right|,\\
    V''(\Phi)&\simeq\frac{M_P^2}{\zeta}\left(2\ln\left|\frac{M_P}{\Phi}\right|+\ln\zeta-3\right).
\end{align}
The second derivative is still divergent, but the degree of divergence has reduced to the logarithmic level. This means the invalid interval of $\Phi$ where $m_\Phi>M_P$ becomes extremely narrow for sufficiently large $\zeta$. Specifically, it can be estimated to $|\Phi|<e^{-\zeta/2}M_P$ for $\zeta\gg1$. In fact, $\zeta$ allows for a very large value up to $10^{10}$ (shown in the next Section), thus the premise of our discussion is satisfied.

Since $\Phi$ represents only the classical background field and non-negligible quantum fluctuations $\delta\phi$ are always present, one can expect that the characteristic scale of $\delta\phi$ is far larger than the narrow classically invalid interval derived above. In other words, the interval where the classical analysis breaks down can be shielded by quantum effects. Therefore, the exponential-extended $\hat R^2$ models significantly alleviate the divergence problem inherent in the polynomial-extended models.

\begin{figure}[!t]
    \centering\includegraphics[width=0.5\textwidth]{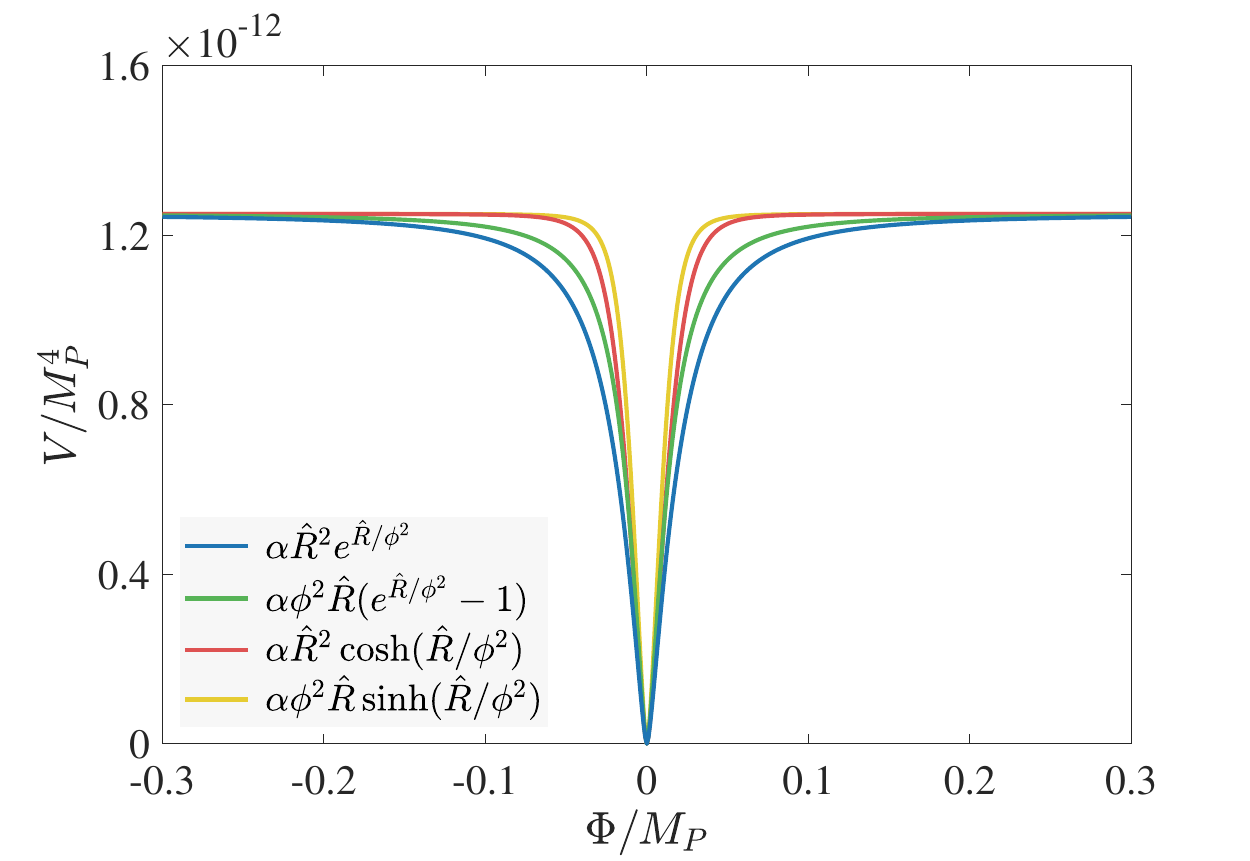}
    \caption{Effective potentials of Weyl exponential-extended models with $\alpha=10^{11}$, $\zeta=10^8$ as an example. \label{fig1}}
\end{figure}

By solving Eq.~(\ref{fixgauge}) numerically, the effective scalar potentials of the above exponential-extended models are depicted in Fig.~\ref{fig1}. As expected, the $\hat R^2$ term dominates for $\Phi\to\infty$, inducing a flat potential with scale-invariant scalar spectrum. As $\Phi$ evolves, the higher-order curvatures become significant, leading to a gradual reduction of $n_s$ from 1. The parameter $\zeta$ controls the width of the pit in the potential: larger values of $\zeta$ result in a wider pit, while smaller values lead to a narrower one. Actually, almost any higher-order curvature extension is capable of inducing potentials like this. The exponential extensions are only illustrative examples. Provided that the coefficients of higher-order terms in the expanded model are significantly larger than those of the relatively lower-order terms, the model can suppress the mass divergence of the inflaton and yield effects analogous to those in our example models. Thus, this inflationary scenario is not a specific construction. It allows a much broader class of realizations.

{\it Observables} —
Now we calculate the inflationary observables, mainly the spectral index $n_s$ and the tensor-to-scalar ratio $r$. They are expressed as
\begin{align}
    n_s=1-6\epsilon_V+2\eta_V,~r=16\epsilon_V,
\end{align}
where the slow-roll parameters
\begin{align}
    \epsilon_V\equiv\frac{M_P^2}{2}\left[\frac{V'(\Phi)}{V(\Phi)}\right]^2,~\eta_V\equiv M_P^2\frac{V''(\Phi)}{V(\Phi)}.
\end{align}
We first present an analytical approach. Although these exponential-extended models lack an analytical expression for $V(\phi)$, they can be approximated by the polynomial model within the slow-roll region. We have classified these models into type A and type B in the previous Section, respectively corresponding to the $n=3$ and $n=4$ cases of model (\ref{Fpoly}). Both cases admit explicit expressions for $V(\phi)$, allowing for analytical treatment. 

We elaborate in detail on the type-A model as an illustrative example, which is $F(\hat R,\phi)=\alpha\hat R^2+\lambda_3\hat R^3/\phi^2$ to the sub-leading-order. Solving Eq.~(\ref{fixgauge}), we have
\begin{align}
    \chi=\frac{\alpha\phi^2}{3\lambda_3}\left(-1+\sqrt{1+\frac{3\lambda_3M_P^2}{\alpha^2\phi^2}}\right).
\end{align}
For simplicity, the approximation $\phi^2 \sim\Phi^2/\zeta$ is adopted. This is applicable for inflaton variations $\Delta\Phi\lesssim\sqrt{6}M_P$ after horizon crossing, which, within the model under consideration, corresponds to $\zeta\lesssim10^9$. With this approximation and a redefined parameter $\xi\equiv\alpha^2/\left(3\lambda_3\zeta\right)$, the effective potential is derived as
\begin{align}
    V(\Phi)\simeq-\frac{\xi M_P^2\Phi^2}{2\alpha}+\frac{\xi^2\Phi^4}{3\alpha}\left[\left(1+\frac{M_P^2}{\xi\Phi^2}\right)^{3/2}-1\right].
\end{align}
Further defining $u\equiv\sqrt{1+M_P^2/(\xi\Phi^2)}$ for convenience, the slow-roll parameters are expressed as
\begin{align}
    &\epsilon_V=2\xi\frac{\left(u^2-1\right)\left(u-1\right)^2}{\left(2u+1\right)^2},\\
    &\eta_V=-6\xi\frac{\left(u^2-1\right)\left(u-1\right)}{u\left(2u+1\right)},
\end{align}
and the $e$-folding number becomes
\begin{align}
    N&\simeq\int^{\Phi^*}_{\sim0} \frac{\md\Phi}{\sqrt{2\epsilon_V}M_P}\nonumber\\
    &=\frac{1}{2\xi}\left[\frac{3}{16}\ln\left(\frac{u-1}{u+1}\right)+\frac{\left(3u-1\right)\left(u+2\right)}{8\left(u-1\right)^2\left(u+1\right)}\right]^{u^*}_\infty,
\end{align}
where $\Phi^*$ denotes the field value when horizon crossing. For $N\sim50-60$ and $\zeta\lesssim10^9$, actually, the first logarithmic term can be ignored due to its negligible contribution. Further defining $u^*\equiv1+\delta$ with $\delta\ll1$, this equation reduces to $N\simeq\left(6+11\delta\right)/\left(32\xi\delta^2\right)$, and we obtain the following approximate solution,
\begin{align}
    u^*\simeq1+\frac{11+\sqrt{121+768\xi N}}{64\xi N}.
\end{align}
Substituting it into the slow-roll parameters and expanding to the sub-leading-order, the inflationary observables are finally derived as
\begin{align}
    n_s&\simeq1-\frac{3}{2N}-\frac{3\sqrt3}{8\sqrt\xi N^{3/2}},\\
    r&\simeq\frac{1}{\sqrt{3\xi} N^{3/2}}+\frac{23}{48\xi N^2}.
\end{align}

Clearly, for the type-A model, the spectral index reads $n_s\simeq1-3/(2N)\sim0.97-0.975$ at the leading order. The type-B model can also be treated analytically in principle. However, it requires massive approximations and cumbersome steps. Skipping tedious calculations, we give a leading order result: $n_s\simeq1-5/(3N)\sim0.967-0.972$. As a reminder, the ACT/SPT $2\sigma$ bounds lies in $n_s\sim0.967-0.98$. Thus, a minor higher-order correction to the $\hat R^2$ model suffices to tilt the scalar spectrum from exact scale invariance to a level that closely matches the latest constraints.

\begin{figure}[!t]
    \centering\includegraphics[width=0.5\textwidth]{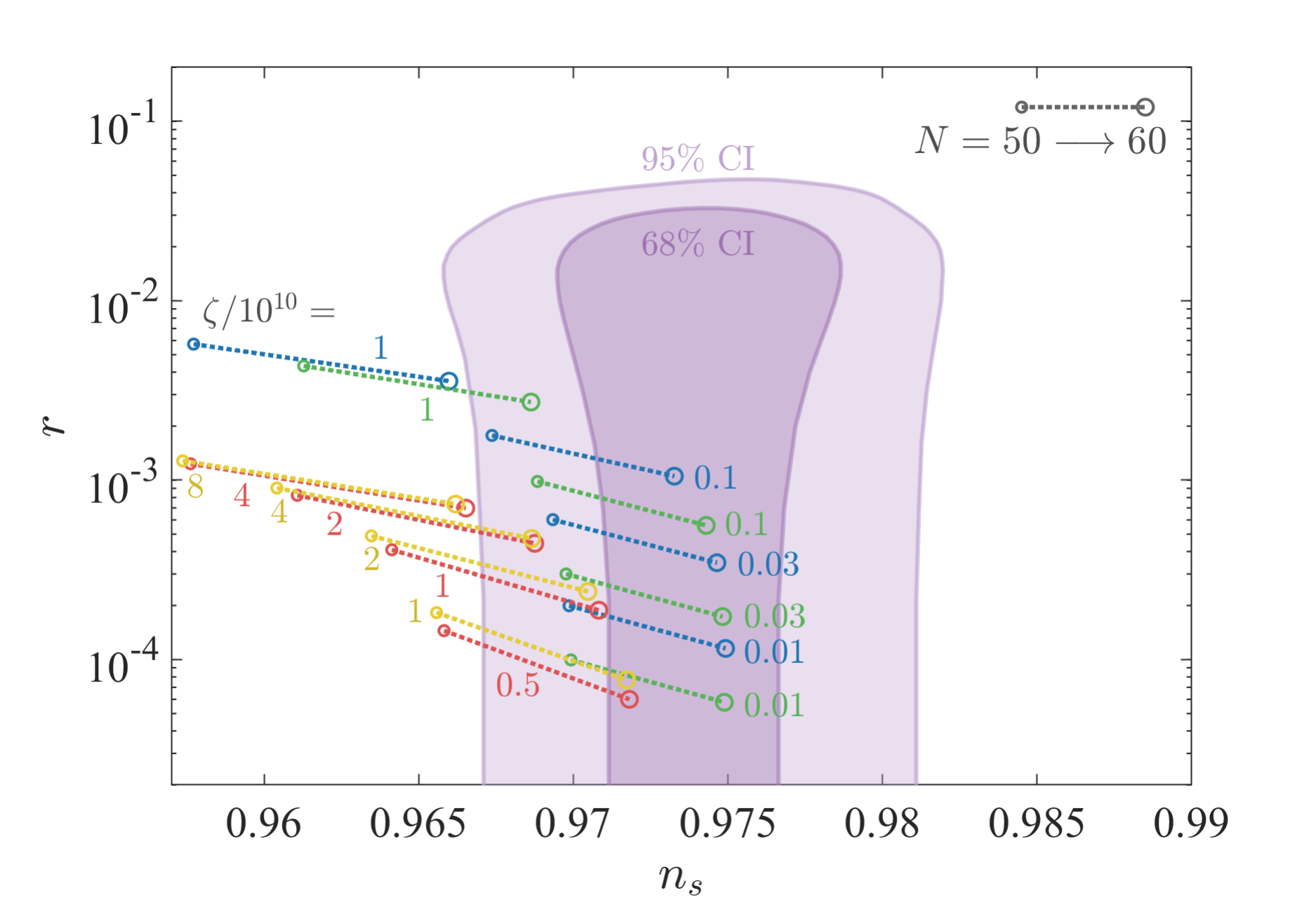}
    \caption{Predictions of spectral index $n_s$ combined with tensor-to-scalar ratio $r$ in the Weyl exponential-extended models with various $\zeta$. The blue, green, red, and yellow dotted lines are respectively for models (\ref{modele}), (\ref{modele-1}), (\ref{modelch}), and (\ref{modelsh}). The violet area is the latest observational constraints given by the ACT collaboration \cite{AtacamaCosmologyTelescope:2025nti}. \label{fig2}}
\end{figure}

We now present the numerical results, as shown in Fig.~\ref{fig2}. Evidently, the theoretical predictions are in good agreement with the ACT constraints for small $\zeta$, specifically, for $\zeta\lesssim10^9$ in type-A models, and for $\zeta\lesssim10^{10}$ in type-B models. Actually, a small $\zeta$ should be favored because it enable the sub-Planck evolution of the inflaton in the Einstein frame. As $\zeta$ decreases, the predicted $n_s$ converges to a fixed range, matching the leading-order analytical results derived previously. Moreover, the tensor-to-scalar ratio is strongly depressed, scaling as $r\propto\zeta$ and $r\propto\zeta^2$ for type-A and type-B models, respectively.

It is necessary to discuss the potential impact of other possible correction terms on the observables. From the symmetry perspective, one might expect additional scale-invariant terms to appear in the model, such as $\phi^2\hat R$, $\phi^4$, $\phi^{2n+4}/\hat R^n$, or terms composed of the Ricci or Riemann tensors. In practice, however, considerations of unitarity and stability preclude terms involving the Ricci or Riemann tensors with arbitrary coefficients, due to the appearance of ghost degrees of freedom. The $\phi^{2n+4}/\hat R^n$ terms with $n\geq1$ are likewise disallowed, as they can render the potential negative in the large $\Phi$ region. The only remaining terms permitted by both symmetry and stability are $\phi^2\hat R$ and $\phi^4$, which would modify the potential at large $\Phi$ and consequently affect the prediction of $n_s$. To ensure such modifications remain negligible, the coefficients of these terms must be sufficiently small.

We admit that requiring these terms to be small may appear somewhat {\it ad hoc}. However, this is not a shortcoming unique to our model. Such a hierarchy problem reflects an ubiquitous challenge in inflation model building, often referred to as the eta problem. In the majority of effective field theory realizations of inflation, higher-order operators generally spoil the flatness of the potential unless their coefficients are suppressed. This issue arises in a wide class of models. Thus, while our model shares this limitation, it does so within a context where the negligible corrections are commonly required.

{\it Discussion} —
A comparison with conventional $f(R)$ theories is instructive. In typical $f(R)$ inflation, the higher-order extensions are treated as corrections to the Starobinsky model \cite{Huang:2013hsb, Asaka:2015vza, Cheong:2020rao, Rodrigues-da-Silva:2021jab, Ivanov:2021chn, Shtanov:2022pdx, Modak:2022gol,Gialamas:2025ofz,Odintsov:2025eiv,Qiu:2025iqm,Bezerra-Sobrinho:2025gfg}. From the perspective of phenomenology, the Starobinsky prediction, $n_s\simeq1-2/N\sim0.96-0.967$, lies slightly below the ACT/SPT preferred range. Positive higher-order curvature corrections would further diminish $n_s$, exacerbating the tension with observations. Conversely, negative corrections could raise $n_s$ into agreement. However, the parameter fine-tuning is required to some extent. Our models are in a different situation. The leading-order result of $n_s$ conforms directly to the ACT/SPT constraints, thereby avoiding fine adjustment of parameters to fit the observations.

Moreover, from the perspective of theoretical completeness, inflation in the $f(R)$ framework with negative corrections no longer originates from a de Sitter state with constant curvature. The Ricci scalar $R$ is not convergent toward the beginning of inflation, and a critical value of $R_c$ (or corresponding $\Phi_c$ in the Einstein frame) emerges where $f(R)$ or $f''(R)$ turns negative. This necessitates a cutoff at $R_c$, rendering the theory UV incomplete \cite{Huang:2013hsb,Qiu:2025iqm}. In contrast, our models preserve the UV completeness, which can be verified that $R_{\phi\to\infty}=M_P^2/(2\alpha)$, corresponding to a constant-curvature origin.

We proceed to a brief discussion of the dark matter candidate in our models, the Weyl gauge boson $\overline w_\mu$. The production mechanism of this particle in the Weyl $\phi^2\hat R+\hat R^2$ model with a Higgs-like potential [Eq.~(\ref{VWR2})] has been investigated in detail in our previous work \cite{Wang:2022ojc}. The inflation models proposed in this work, however, lead to a different outcome. First, the mass $m_W$ of $\overline w_\mu$ depends on the inflaton $\Phi$ during inflation [see Eq.~(\ref{LEWeyl})], affecting the evolution of its longitudinal mode. In the Higgs-like potential model, the increase of $|\Phi|$ and $m_W$ during inflation provides negative contributions to the frequency term $\omega_k^2$ of the longitudinal mode, thus promoting the gravitational production of $\overline w_\mu$. By contrast, for models in this work, $|\Phi|$ and $m_W$ decrease during inflation, leading to a suppression of the gravitational production. We estimate that the yield can be reduced by two orders of magnitude compared to the previous scenario.

Meanwhile, these models are able to realize small-field inflation, which significantly lowers the Hubble parameter $H_e$ at the end of inflation and further suppresses the dark matter yield. Given that the relic energy fraction of gravitationally produced vector dark matter (for $m\ll H_\mathrm{reh}$) follows $\Omega_{\mathrm{DM}}\propto H_e^2m^{1/2}$ \cite{Graham:2015rva,Ema:2019yrd,Ahmed:2020fhc}, we estimate that the minimum mass of $\overline w_\mu$ required to match the observed $\Omega_{\mathrm{DM}}\sim0.26$ increases by many orders of magnitude—from $10^{-4}$ eV reported in \cite{Wang:2022ojc} up to the eV or even GeV scale.

Moreover, in the $\phi^2\hat R+\hat R^2$ model, the coupling of $\Phi$ and $\overline w_\mu$ generates $\Phi\overline w_\mu\overline w^\mu$ at the leading order due to the nonzero vacuum expectation value of $\Phi$. This term allows the inflaton to decay spontaneously into $\overline w_\mu$, which would result in excessive dark radiation if the reheating temperature is  relatively low. In contrast, for models in this work, the leading-order coupling takes the form $\Phi^2\overline w_\mu\overline w^\mu$, which forbids such spontaneous decays, thereby avoiding the overproduction of dark radiation.

{\it Conclusion} —
In this work, we demonstrate a viable scenario that inflation arises from the higher-order curvature extensions, especially the divergence-suppressed extensions with exponential forms, on the basis of the Weyl scale-invariant $\hat R^2$ model. In these models, the $\hat R^2$-induced cosmological constant serves as the origin of scale invariance for the scalar perturbation spectrum, while the higher-order curvatures bring about the observed slight deviation. Remarkably, such deviations align perfectly with the latest CMB observations across a broad parameter range. At the leading order, the spectral index is given by $n_s\simeq1-3/(2N)$ or $n_s\simeq1-5/(3N)$ for various models, which lies comfortably within the 95\% confidence interval reported by ACT/SPT. Consequently, a scale-invariant inflationary framework naturally explains the observed nearly scale-invariant spectrum, implying an enduring cosmic echo of the primordial symmetry.

\bibliography{ref.bib}

%apsrev4-2.bst 2019-01-14 (MD) hand-edited version of apsrev4-1.bst
%Control: key (0)
%Control: author (8) initials jnrlst
%Control: editor formatted (1) identically to author
%Control: production of article title (0) allowed
%Control: page (0) single
%Control: year (1) truncated
%Control: production of eprint (0) enabled
\begin{thebibliography}{62}%
\makeatletter
\providecommand \@ifxundefined [1]{%
 \@ifx{#1\undefined}
}%
\providecommand \@ifnum [1]{%
 \ifnum #1\expandafter \@firstoftwo
 \else \expandafter \@secondoftwo
 \fi
}%
\providecommand \@ifx [1]{%
 \ifx #1\expandafter \@firstoftwo
 \else \expandafter \@secondoftwo
 \fi
}%
\providecommand \natexlab [1]{#1}%
\providecommand \enquote  [1]{``#1''}%
\providecommand \bibnamefont  [1]{#1}%
\providecommand \bibfnamefont [1]{#1}%
\providecommand \citenamefont [1]{#1}%
\providecommand \href@noop [0]{\@secondoftwo}%
\providecommand \href [0]{\begingroup \@sanitize@url \@href}%
\providecommand \@href[1]{\@@startlink{#1}\@@href}%
\providecommand \@@href[1]{\endgroup#1\@@endlink}%
\providecommand \@sanitize@url [0]{\catcode `\\12\catcode `\$12\catcode
  `\&12\catcode `\#12\catcode `\^12\catcode `\_12\catcode `\%12\relax}%
\providecommand \@@startlink[1]{}%
\providecommand \@@endlink[0]{}%
\providecommand \url  [0]{\begingroup\@sanitize@url \@url }%
\providecommand \@url [1]{\endgroup\@href {#1}{\urlprefix }}%
\providecommand \urlprefix  [0]{URL }%
\providecommand \Eprint [0]{\href }%
\providecommand \doibase [0]{https://doi.org/}%
\providecommand \selectlanguage [0]{\@gobble}%
\providecommand \bibinfo  [0]{\@secondoftwo}%
\providecommand \bibfield  [0]{\@secondoftwo}%
\providecommand \translation [1]{[#1]}%
\providecommand \BibitemOpen [0]{}%
\providecommand \bibitemStop [0]{}%
\providecommand \bibitemNoStop [0]{.\EOS\space}%
\providecommand \EOS [0]{\spacefactor3000\relax}%
\providecommand \BibitemShut  [1]{\csname bibitem#1\endcsname}%
\let\auto@bib@innerbib\@empty
%</preamble>
\bibitem [{\citenamefont {Louis}\ \emph {et~al.}(2025)\citenamefont {Louis}
  \emph {et~al.}}]{AtacamaCosmologyTelescope:2025blo}%
  \BibitemOpen
  \bibfield  {author} {\bibinfo {author} {\bibfnamefont {T.}~\bibnamefont
  {Louis}} \emph {et~al.} (\bibinfo {collaboration} {Atacama Cosmology
  Telescope}),\ }\bibfield  {title} {\bibinfo {title} {{The Atacama Cosmology
  Telescope: DR6 power spectra, likelihoods and {\ensuremath{\Lambda}}CDM
  parameters}},\ }\href {https://doi.org/10.1088/1475-7516/2025/11/062}
  {\bibfield  {journal} {\bibinfo  {journal} {JCAP}\ }\textbf {\bibinfo
  {volume} {11}},\ \bibinfo {pages} {062}},\ \Eprint
  {https://arxiv.org/abs/2503.14452} {arXiv:2503.14452 [astro-ph.CO]}
  \BibitemShut {NoStop}%
\bibitem [{\citenamefont {Calabrese}\ \emph {et~al.}(2025)\citenamefont
  {Calabrese} \emph {et~al.}}]{AtacamaCosmologyTelescope:2025nti}%
  \BibitemOpen
  \bibfield  {author} {\bibinfo {author} {\bibfnamefont {E.}~\bibnamefont
  {Calabrese}} \emph {et~al.} (\bibinfo {collaboration} {Atacama Cosmology
  Telescope}),\ }\bibfield  {title} {\bibinfo {title} {{The Atacama Cosmology
  Telescope: DR6 constraints on extended cosmological models}},\ }\href
  {https://doi.org/10.1088/1475-7516/2025/11/063} {\bibfield  {journal}
  {\bibinfo  {journal} {JCAP}\ }\textbf {\bibinfo {volume} {11}},\ \bibinfo
  {pages} {063}},\ \Eprint {https://arxiv.org/abs/2503.14454} {arXiv:2503.14454
  [astro-ph.CO]} \BibitemShut {NoStop}%
\bibitem [{\citenamefont {Camphuis}\ \emph {et~al.}(2026)\citenamefont
  {Camphuis} \emph {et~al.}}]{SPT-3G:2025bzu}%
  \BibitemOpen
  \bibfield  {author} {\bibinfo {author} {\bibfnamefont {E.}~\bibnamefont
  {Camphuis}} \emph {et~al.} (\bibinfo {collaboration} {SPT-3G}),\ }\bibfield
  {title} {\bibinfo {title} {{SPT-3G D1: CMB temperature and polarization power
  spectra and cosmology from 2019 and 2020 observations of the SPT-3G main
  field}},\ }\href {https://doi.org/10.1103/7wt3-9v2y} {\bibfield  {journal}
  {\bibinfo  {journal} {Phys. Rev. D}\ }\textbf {\bibinfo {volume} {113}},\
  \bibinfo {pages} {083504} (\bibinfo {year} {2026})},\ \Eprint
  {https://arxiv.org/abs/2506.20707} {arXiv:2506.20707 [astro-ph.CO]}
  \BibitemShut {NoStop}%
\bibitem [{\citenamefont {Aghanim}\ \emph {et~al.}(2020)\citenamefont {Aghanim}
  \emph {et~al.}}]{Planck:2018vyg}%
  \BibitemOpen
  \bibfield  {author} {\bibinfo {author} {\bibfnamefont {N.}~\bibnamefont
  {Aghanim}} \emph {et~al.} (\bibinfo {collaboration} {Planck}),\ }\bibfield
  {title} {\bibinfo {title} {{Planck 2018 results. VI. Cosmological
  parameters}},\ }\href {https://doi.org/10.1051/0004-6361/201833910}
  {\bibfield  {journal} {\bibinfo  {journal} {Astron. Astrophys.}\ }\textbf
  {\bibinfo {volume} {641}},\ \bibinfo {pages} {A6} (\bibinfo {year} {2020})},\
  \bibinfo {note} {[Erratum: Astron.Astrophys. 652, C4 (2021)]},\ \Eprint
  {https://arxiv.org/abs/1807.06209} {arXiv:1807.06209 [astro-ph.CO]}
  \BibitemShut {NoStop}%
\bibitem [{\citenamefont {Abdul~Karim}\ \emph {et~al.}(2025)\citenamefont
  {Abdul~Karim} \emph {et~al.}}]{DESI:2025zgx}%
  \BibitemOpen
  \bibfield  {author} {\bibinfo {author} {\bibfnamefont {M.}~\bibnamefont
  {Abdul~Karim}} \emph {et~al.} (\bibinfo {collaboration} {DESI}),\ }\bibfield
  {title} {\bibinfo {title} {{DESI DR2 results. II. Measurements of baryon
  acoustic oscillations and cosmological constraints}},\ }\href
  {https://doi.org/10.1103/tr6y-kpc6} {\bibfield  {journal} {\bibinfo
  {journal} {Phys. Rev. D}\ }\textbf {\bibinfo {volume} {112}},\ \bibinfo
  {pages} {083515} (\bibinfo {year} {2025})},\ \Eprint
  {https://arxiv.org/abs/2503.14738} {arXiv:2503.14738 [astro-ph.CO]}
  \BibitemShut {NoStop}%
\bibitem [{\citenamefont {Mukhanov}\ and\ \citenamefont
  {Chibisov}(1981)}]{Mukhanov:1981xt}%
  \BibitemOpen
  \bibfield  {author} {\bibinfo {author} {\bibfnamefont {V.~F.}\ \bibnamefont
  {Mukhanov}}\ and\ \bibinfo {author} {\bibfnamefont {G.~V.}\ \bibnamefont
  {Chibisov}},\ }\bibfield  {title} {\bibinfo {title} {{Quantum Fluctuations
  and a Nonsingular Universe}},\ }\href@noop {} {\bibfield  {journal} {\bibinfo
   {journal} {JETP Lett.}\ }\textbf {\bibinfo {volume} {33}},\ \bibinfo {pages}
  {532} (\bibinfo {year} {1981})}\BibitemShut {NoStop}%
\bibitem [{\citenamefont {Starobinsky}(1980)}]{Starobinsky:1980te}%
  \BibitemOpen
  \bibfield  {author} {\bibinfo {author} {\bibfnamefont {A.~A.}\ \bibnamefont
  {Starobinsky}},\ }\bibfield  {title} {\bibinfo {title} {{A New Type of
  Isotropic Cosmological Models Without Singularity}},\ }\href
  {https://doi.org/10.1016/0370-2693(80)90670-X} {\bibfield  {journal}
  {\bibinfo  {journal} {Phys. Lett. B}\ }\textbf {\bibinfo {volume} {91}},\
  \bibinfo {pages} {99} (\bibinfo {year} {1980})}\BibitemShut {NoStop}%
\bibitem [{\citenamefont {Bezrukov}\ and\ \citenamefont
  {Shaposhnikov}(2008)}]{Bezrukov:2007ep}%
  \BibitemOpen
  \bibfield  {author} {\bibinfo {author} {\bibfnamefont {F.~L.}\ \bibnamefont
  {Bezrukov}}\ and\ \bibinfo {author} {\bibfnamefont {M.}~\bibnamefont
  {Shaposhnikov}},\ }\bibfield  {title} {\bibinfo {title} {{The Standard Model
  Higgs boson as the inflaton}},\ }\href
  {https://doi.org/10.1016/j.physletb.2007.11.072} {\bibfield  {journal}
  {\bibinfo  {journal} {Phys. Lett. B}\ }\textbf {\bibinfo {volume} {659}},\
  \bibinfo {pages} {703} (\bibinfo {year} {2008})},\ \Eprint
  {https://arxiv.org/abs/0710.3755} {arXiv:0710.3755 [hep-th]} \BibitemShut
  {NoStop}%
\bibitem [{\citenamefont {Kallosh}\ \emph {et~al.}(2013)\citenamefont
  {Kallosh}, \citenamefont {Linde},\ and\ \citenamefont
  {Roest}}]{Kallosh:2013yoa}%
  \BibitemOpen
  \bibfield  {author} {\bibinfo {author} {\bibfnamefont {R.}~\bibnamefont
  {Kallosh}}, \bibinfo {author} {\bibfnamefont {A.}~\bibnamefont {Linde}},\
  and\ \bibinfo {author} {\bibfnamefont {D.}~\bibnamefont {Roest}},\ }\bibfield
   {title} {\bibinfo {title} {{Superconformal Inflationary
  $\alpha$-Attractors}},\ }\href {https://doi.org/10.1007/JHEP11(2013)198}
  {\bibfield  {journal} {\bibinfo  {journal} {JHEP}\ }\textbf {\bibinfo
  {volume} {11}},\ \bibinfo {pages} {198}},\ \Eprint
  {https://arxiv.org/abs/1311.0472} {arXiv:1311.0472 [hep-th]} \BibitemShut
  {NoStop}%
\bibitem [{\citenamefont {Weyl}(1918)}]{Weyl:1918ib}%
  \BibitemOpen
  \bibfield  {author} {\bibinfo {author} {\bibfnamefont {H.}~\bibnamefont
  {Weyl}},\ }\bibfield  {title} {\bibinfo {title} {{Gravitation and
  electricity}},\ }\href@noop {} {\bibfield  {journal} {\bibinfo  {journal}
  {Sitzungsber. Preuss. Akad. Wiss. Berlin (Math. Phys. )}\ }\textbf {\bibinfo
  {volume} {1918}},\ \bibinfo {pages} {465} (\bibinfo {year}
  {1918})}\BibitemShut {NoStop}%
\bibitem [{\citenamefont {Weyl}(1919)}]{Weyl:1919fi}%
  \BibitemOpen
  \bibfield  {author} {\bibinfo {author} {\bibfnamefont {H.}~\bibnamefont
  {Weyl}},\ }\bibfield  {title} {\bibinfo {title} {{A New Extension of
  Relativity Theory}},\ }\href {https://doi.org/10.1002/andp.19193641002}
  {\bibfield  {journal} {\bibinfo  {journal} {Annalen Phys.}\ }\textbf
  {\bibinfo {volume} {59}},\ \bibinfo {pages} {101} (\bibinfo {year}
  {1919})}\BibitemShut {NoStop}%
\bibitem [{\citenamefont {Cheng}(1988)}]{Cheng:1988zx}%
  \BibitemOpen
  \bibfield  {author} {\bibinfo {author} {\bibfnamefont {H.}~\bibnamefont
  {Cheng}},\ }\bibfield  {title} {\bibinfo {title} {{The Possible Existence of
  Weyl's Vector Meson}},\ }\href {https://doi.org/10.1103/PhysRevLett.61.2182}
  {\bibfield  {journal} {\bibinfo  {journal} {Phys. Rev. Lett.}\ }\textbf
  {\bibinfo {volume} {61}},\ \bibinfo {pages} {2182} (\bibinfo {year}
  {1988})}\BibitemShut {NoStop}%
\bibitem [{\citenamefont {Scholz}(2015)}]{Scholz:2014kba}%
  \BibitemOpen
  \bibfield  {author} {\bibinfo {author} {\bibfnamefont {E.}~\bibnamefont
  {Scholz}},\ }\bibfield  {title} {\bibinfo {title} {{Higgs and gravitational
  scalar fields together induce Weyl gauge}},\ }\href
  {https://doi.org/10.1007/s10714-015-1854-z} {\bibfield  {journal} {\bibinfo
  {journal} {Gen. Rel. Grav.}\ }\textbf {\bibinfo {volume} {47}},\ \bibinfo
  {pages} {7} (\bibinfo {year} {2015})},\ \Eprint
  {https://arxiv.org/abs/1407.6811} {arXiv:1407.6811 [gr-qc]} \BibitemShut
  {NoStop}%
\bibitem [{\citenamefont {Ohanian}(2016)}]{Ohanian:2015wva}%
  \BibitemOpen
  \bibfield  {author} {\bibinfo {author} {\bibfnamefont {H.~C.}\ \bibnamefont
  {Ohanian}},\ }\bibfield  {title} {\bibinfo {title} {{Weyl gauge-vector and
  complex dilaton scalar for conformal symmetry and its breaking}},\ }\href
  {https://doi.org/10.1007/s10714-016-2023-8} {\bibfield  {journal} {\bibinfo
  {journal} {Gen. Rel. Grav.}\ }\textbf {\bibinfo {volume} {48}},\ \bibinfo
  {pages} {25} (\bibinfo {year} {2016})},\ \Eprint
  {https://arxiv.org/abs/1502.00020} {arXiv:1502.00020 [gr-qc]} \BibitemShut
  {NoStop}%
\bibitem [{\citenamefont {Ferreira}\ \emph
  {et~al.}(2018{\natexlab{a}})\citenamefont {Ferreira}, \citenamefont {Hill},\
  and\ \citenamefont {Ross}}]{Ferreira:2018itt}%
  \BibitemOpen
  \bibfield  {author} {\bibinfo {author} {\bibfnamefont {P.~G.}\ \bibnamefont
  {Ferreira}}, \bibinfo {author} {\bibfnamefont {C.~T.}\ \bibnamefont {Hill}},\
  and\ \bibinfo {author} {\bibfnamefont {G.~G.}\ \bibnamefont {Ross}},\
  }\bibfield  {title} {\bibinfo {title} {{Inertial Spontaneous Symmetry
  Breaking and Quantum Scale Invariance}},\ }\href
  {https://doi.org/10.1103/PhysRevD.98.116012} {\bibfield  {journal} {\bibinfo
  {journal} {Phys. Rev. D}\ }\textbf {\bibinfo {volume} {98}},\ \bibinfo
  {pages} {116012} (\bibinfo {year} {2018}{\natexlab{a}})},\ \Eprint
  {https://arxiv.org/abs/1801.07676} {arXiv:1801.07676 [hep-th]} \BibitemShut
  {NoStop}%
\bibitem [{\citenamefont {Ghilencea}(2019{\natexlab{a}})}]{Ghilencea:2018dqd}%
  \BibitemOpen
  \bibfield  {author} {\bibinfo {author} {\bibfnamefont {D.~M.}\ \bibnamefont
  {Ghilencea}},\ }\bibfield  {title} {\bibinfo {title} {{Spontaneous breaking
  of Weyl quadratic gravity to Einstein action and Higgs potential}},\ }\href
  {https://doi.org/10.1007/JHEP03(2019)049} {\bibfield  {journal} {\bibinfo
  {journal} {JHEP}\ }\textbf {\bibinfo {volume} {03}},\ \bibinfo {pages}
  {049}},\ \Eprint {https://arxiv.org/abs/1812.08613} {arXiv:1812.08613
  [hep-th]} \BibitemShut {NoStop}%
\bibitem [{\citenamefont {Oda}(2020)}]{Oda:2020cmi}%
  \BibitemOpen
  \bibfield  {author} {\bibinfo {author} {\bibfnamefont {I.}~\bibnamefont
  {Oda}},\ }\bibfield  {title} {\bibinfo {title} {{Higgs Potential from Weyl
  Conformal Gravity}},\ }\href {https://doi.org/10.1142/S0217732320503046}
  {\bibfield  {journal} {\bibinfo  {journal} {Mod. Phys. Lett. A}\ }\textbf
  {\bibinfo {volume} {35}},\ \bibinfo {pages} {2050304} (\bibinfo {year}
  {2020})},\ \Eprint {https://arxiv.org/abs/2006.10867} {arXiv:2006.10867
  [hep-th]} \BibitemShut {NoStop}%
\bibitem [{\citenamefont {Ghilencea}(2022)}]{Ghilencea:2021lpa}%
  \BibitemOpen
  \bibfield  {author} {\bibinfo {author} {\bibfnamefont {D.~M.}\ \bibnamefont
  {Ghilencea}},\ }\bibfield  {title} {\bibinfo {title} {{Standard Model in Weyl
  conformal geometry}},\ }\href
  {https://doi.org/10.1140/epjc/s10052-021-09887-y} {\bibfield  {journal}
  {\bibinfo  {journal} {Eur. Phys. J. C}\ }\textbf {\bibinfo {volume} {82}},\
  \bibinfo {pages} {23} (\bibinfo {year} {2022})},\ \Eprint
  {https://arxiv.org/abs/2104.15118} {arXiv:2104.15118 [hep-ph]} \BibitemShut
  {NoStop}%
\bibitem [{\citenamefont {Ferreira}\ \emph {et~al.}(2019)\citenamefont
  {Ferreira}, \citenamefont {Hill}, \citenamefont {Noller},\ and\ \citenamefont
  {Ross}}]{Ferreira:2019zzx}%
  \BibitemOpen
  \bibfield  {author} {\bibinfo {author} {\bibfnamefont {P.~G.}\ \bibnamefont
  {Ferreira}}, \bibinfo {author} {\bibfnamefont {C.~T.}\ \bibnamefont {Hill}},
  \bibinfo {author} {\bibfnamefont {J.}~\bibnamefont {Noller}},\ and\ \bibinfo
  {author} {\bibfnamefont {G.~G.}\ \bibnamefont {Ross}},\ }\bibfield  {title}
  {\bibinfo {title} {{Scale-independent $R^2$ inflation}},\ }\href
  {https://doi.org/10.1103/PhysRevD.100.123516} {\bibfield  {journal} {\bibinfo
   {journal} {Phys. Rev. D}\ }\textbf {\bibinfo {volume} {100}},\ \bibinfo
  {pages} {123516} (\bibinfo {year} {2019})},\ \Eprint
  {https://arxiv.org/abs/1906.03415} {arXiv:1906.03415 [gr-qc]} \BibitemShut
  {NoStop}%
\bibitem [{\citenamefont {Ghilencea}(2019{\natexlab{b}})}]{Ghilencea:2019rqj}%
  \BibitemOpen
  \bibfield  {author} {\bibinfo {author} {\bibfnamefont {D.~M.}\ \bibnamefont
  {Ghilencea}},\ }\bibfield  {title} {\bibinfo {title} {{Weyl R$^{2}$ inflation
  with an emergent Planck scale}},\ }\href
  {https://doi.org/10.1007/JHEP10(2019)209} {\bibfield  {journal} {\bibinfo
  {journal} {JHEP}\ }\textbf {\bibinfo {volume} {10}},\ \bibinfo {pages}
  {209}},\ \Eprint {https://arxiv.org/abs/1906.11572} {arXiv:1906.11572
  [gr-qc]} \BibitemShut {NoStop}%
\bibitem [{\citenamefont {Ghilencea}(2020)}]{Ghilencea:2020piz}%
  \BibitemOpen
  \bibfield  {author} {\bibinfo {author} {\bibfnamefont {D.~M.}\ \bibnamefont
  {Ghilencea}},\ }\bibfield  {title} {\bibinfo {title} {{Palatini quadratic
  gravity: spontaneous breaking of gauged scale symmetry and inflation}},\
  }\href {https://doi.org/10.1140/epjc/s10052-020-08722-0} {\bibfield
  {journal} {\bibinfo  {journal} {Eur. Phys. J. C}\ }\textbf {\bibinfo {volume}
  {80}},\ \bibinfo {pages} {1147} (\bibinfo {year} {2020})},\ \Eprint
  {https://arxiv.org/abs/2003.08516} {arXiv:2003.08516 [hep-th]} \BibitemShut
  {NoStop}%
\bibitem [{\citenamefont {Tang}\ and\ \citenamefont
  {Wu}(2020{\natexlab{a}})}]{Tang:2020ovf}%
  \BibitemOpen
  \bibfield  {author} {\bibinfo {author} {\bibfnamefont {Y.}~\bibnamefont
  {Tang}}\ and\ \bibinfo {author} {\bibfnamefont {Y.-L.}\ \bibnamefont {Wu}},\
  }\bibfield  {title} {\bibinfo {title} {{Weyl scaling invariant $R^2$ gravity
  for inflation and dark matter}},\ }\href
  {https://doi.org/10.1016/j.physletb.2020.135716} {\bibfield  {journal}
  {\bibinfo  {journal} {Phys. Lett. B}\ }\textbf {\bibinfo {volume} {809}},\
  \bibinfo {pages} {135716} (\bibinfo {year} {2020}{\natexlab{a}})},\ \Eprint
  {https://arxiv.org/abs/2006.02811} {arXiv:2006.02811 [hep-ph]} \BibitemShut
  {NoStop}%
\bibitem [{\citenamefont {Cai}\ \emph {et~al.}(2022)\citenamefont {Cai},
  \citenamefont {Hao},\ and\ \citenamefont {Wang}}]{Cai:2021png}%
  \BibitemOpen
  \bibfield  {author} {\bibinfo {author} {\bibfnamefont {R.-G.}\ \bibnamefont
  {Cai}}, \bibinfo {author} {\bibfnamefont {Y.-S.}\ \bibnamefont {Hao}},\ and\
  \bibinfo {author} {\bibfnamefont {S.-J.}\ \bibnamefont {Wang}},\ }\bibfield
  {title} {\bibinfo {title} {{Cosmic inflation from broken conformal
  symmetry}},\ }\href {https://doi.org/10.1088/1572-9494/ac6b2f} {\bibfield
  {journal} {\bibinfo  {journal} {Commun. Theor. Phys.}\ }\textbf {\bibinfo
  {volume} {74}},\ \bibinfo {pages} {095401} (\bibinfo {year} {2022})},\
  \Eprint {https://arxiv.org/abs/2110.14718} {arXiv:2110.14718 [gr-qc]}
  \BibitemShut {NoStop}%
\bibitem [{\citenamefont {Smolin}(1979)}]{Smolin:1979uz}%
  \BibitemOpen
  \bibfield  {author} {\bibinfo {author} {\bibfnamefont {L.}~\bibnamefont
  {Smolin}},\ }\bibfield  {title} {\bibinfo {title} {{Towards a Theory of
  Space-Time Structure at Very Short Distances}},\ }\href
  {https://doi.org/10.1016/0550-3213(79)90059-2} {\bibfield  {journal}
  {\bibinfo  {journal} {Nucl. Phys. B}\ }\textbf {\bibinfo {volume} {160}},\
  \bibinfo {pages} {253} (\bibinfo {year} {1979})}\BibitemShut {NoStop}%
\bibitem [{\citenamefont {Nishino}\ and\ \citenamefont
  {Rajpoot}(2009)}]{Nishino:2009in}%
  \BibitemOpen
  \bibfield  {author} {\bibinfo {author} {\bibfnamefont {H.}~\bibnamefont
  {Nishino}}\ and\ \bibinfo {author} {\bibfnamefont {S.}~\bibnamefont
  {Rajpoot}},\ }\bibfield  {title} {\bibinfo {title} {{Implication of
  Compensator Field and Local Scale Invariance in the Standard Model}},\ }\href
  {https://doi.org/10.1103/PhysRevD.79.125025} {\bibfield  {journal} {\bibinfo
  {journal} {Phys. Rev. D}\ }\textbf {\bibinfo {volume} {79}},\ \bibinfo
  {pages} {125025} (\bibinfo {year} {2009})},\ \Eprint
  {https://arxiv.org/abs/0906.4778} {arXiv:0906.4778 [hep-th]} \BibitemShut
  {NoStop}%
\bibitem [{\citenamefont {Bars}\ \emph {et~al.}(2014)\citenamefont {Bars},
  \citenamefont {Steinhardt},\ and\ \citenamefont {Turok}}]{Bars:2013yba}%
  \BibitemOpen
  \bibfield  {author} {\bibinfo {author} {\bibfnamefont {I.}~\bibnamefont
  {Bars}}, \bibinfo {author} {\bibfnamefont {P.}~\bibnamefont {Steinhardt}},\
  and\ \bibinfo {author} {\bibfnamefont {N.}~\bibnamefont {Turok}},\ }\bibfield
   {title} {\bibinfo {title} {{Local Conformal Symmetry in Physics and
  Cosmology}},\ }\href {https://doi.org/10.1103/PhysRevD.89.043515} {\bibfield
  {journal} {\bibinfo  {journal} {Phys. Rev. D}\ }\textbf {\bibinfo {volume}
  {89}},\ \bibinfo {pages} {043515} (\bibinfo {year} {2014})},\ \Eprint
  {https://arxiv.org/abs/1307.1848} {arXiv:1307.1848 [hep-th]} \BibitemShut
  {NoStop}%
\bibitem [{\citenamefont {Quiros}(2014)}]{Quiros:2014hua}%
  \BibitemOpen
  \bibfield  {author} {\bibinfo {author} {\bibfnamefont {I.}~\bibnamefont
  {Quiros}},\ }\bibfield  {title} {\bibinfo {title} {{On the physical
  consequences of a Weyl invariant theory of gravity}},\ }\href@noop {} {\
  (\bibinfo {year} {2014})},\ \Eprint {https://arxiv.org/abs/1401.2643}
  {arXiv:1401.2643 [gr-qc]} \BibitemShut {NoStop}%
\bibitem [{\citenamefont {Ferreira}\ \emph {et~al.}(2017)\citenamefont
  {Ferreira}, \citenamefont {Hill},\ and\ \citenamefont
  {Ross}}]{Ferreira:2016wem}%
  \BibitemOpen
  \bibfield  {author} {\bibinfo {author} {\bibfnamefont {P.~G.}\ \bibnamefont
  {Ferreira}}, \bibinfo {author} {\bibfnamefont {C.~T.}\ \bibnamefont {Hill}},\
  and\ \bibinfo {author} {\bibfnamefont {G.~G.}\ \bibnamefont {Ross}},\
  }\bibfield  {title} {\bibinfo {title} {{Weyl Current, Scale-Invariant
  Inflation and Planck Scale Generation}},\ }\href
  {https://doi.org/10.1103/PhysRevD.95.043507} {\bibfield  {journal} {\bibinfo
  {journal} {Phys. Rev. D}\ }\textbf {\bibinfo {volume} {95}},\ \bibinfo
  {pages} {043507} (\bibinfo {year} {2017})},\ \Eprint
  {https://arxiv.org/abs/1610.09243} {arXiv:1610.09243 [hep-th]} \BibitemShut
  {NoStop}%
\bibitem [{\citenamefont {Ferreira}\ \emph
  {et~al.}(2018{\natexlab{b}})\citenamefont {Ferreira}, \citenamefont {Hill},
  \citenamefont {Noller},\ and\ \citenamefont {Ross}}]{Ferreira:2018qss}%
  \BibitemOpen
  \bibfield  {author} {\bibinfo {author} {\bibfnamefont {P.~G.}\ \bibnamefont
  {Ferreira}}, \bibinfo {author} {\bibfnamefont {C.~T.}\ \bibnamefont {Hill}},
  \bibinfo {author} {\bibfnamefont {J.}~\bibnamefont {Noller}},\ and\ \bibinfo
  {author} {\bibfnamefont {G.~G.}\ \bibnamefont {Ross}},\ }\bibfield  {title}
  {\bibinfo {title} {{Inflation in a scale invariant universe}},\ }\href
  {https://doi.org/10.1103/PhysRevD.97.123516} {\bibfield  {journal} {\bibinfo
  {journal} {Phys. Rev. D}\ }\textbf {\bibinfo {volume} {97}},\ \bibinfo
  {pages} {123516} (\bibinfo {year} {2018}{\natexlab{b}})},\ \Eprint
  {https://arxiv.org/abs/1802.06069} {arXiv:1802.06069 [astro-ph.CO]}
  \BibitemShut {NoStop}%
\bibitem [{\citenamefont {Tang}\ and\ \citenamefont {Wu}(2018)}]{Tang:2018mhn}%
  \BibitemOpen
  \bibfield  {author} {\bibinfo {author} {\bibfnamefont {Y.}~\bibnamefont
  {Tang}}\ and\ \bibinfo {author} {\bibfnamefont {Y.-L.}\ \bibnamefont {Wu}},\
  }\bibfield  {title} {\bibinfo {title} {{Inflation in gauge theory of gravity
  with local scaling symmetry and quantum induced symmetry breaking}},\ }\href
  {https://doi.org/10.1016/j.physletb.2018.07.048} {\bibfield  {journal}
  {\bibinfo  {journal} {Phys. Lett. B}\ }\textbf {\bibinfo {volume} {784}},\
  \bibinfo {pages} {163} (\bibinfo {year} {2018})},\ \Eprint
  {https://arxiv.org/abs/1805.08507} {arXiv:1805.08507 [gr-qc]} \BibitemShut
  {NoStop}%
\bibitem [{\citenamefont {Ghilencea}\ and\ \citenamefont
  {Lee}(2019)}]{Ghilencea:2018thl}%
  \BibitemOpen
  \bibfield  {author} {\bibinfo {author} {\bibfnamefont {D.~M.}\ \bibnamefont
  {Ghilencea}}\ and\ \bibinfo {author} {\bibfnamefont {H.~M.}\ \bibnamefont
  {Lee}},\ }\bibfield  {title} {\bibinfo {title} {{Weyl gauge symmetry and its
  spontaneous breaking in the standard model and inflation}},\ }\href
  {https://doi.org/10.1103/PhysRevD.99.115007} {\bibfield  {journal} {\bibinfo
  {journal} {Phys. Rev. D}\ }\textbf {\bibinfo {volume} {99}},\ \bibinfo
  {pages} {115007} (\bibinfo {year} {2019})},\ \Eprint
  {https://arxiv.org/abs/1809.09174} {arXiv:1809.09174 [hep-th]} \BibitemShut
  {NoStop}%
\bibitem [{\citenamefont {Tang}\ and\ \citenamefont
  {Wu}(2020{\natexlab{b}})}]{Tang:2019uex}%
  \BibitemOpen
  \bibfield  {author} {\bibinfo {author} {\bibfnamefont {Y.}~\bibnamefont
  {Tang}}\ and\ \bibinfo {author} {\bibfnamefont {Y.-L.}\ \bibnamefont {Wu}},\
  }\bibfield  {title} {\bibinfo {title} {{Weyl Symmetry Inspired Inflation and
  Dark Matter}},\ }\href {https://doi.org/10.1016/j.physletb.2020.135320}
  {\bibfield  {journal} {\bibinfo  {journal} {Phys. Lett. B}\ }\textbf
  {\bibinfo {volume} {803}},\ \bibinfo {pages} {135320} (\bibinfo {year}
  {2020}{\natexlab{b}})},\ \Eprint {https://arxiv.org/abs/1904.04493}
  {arXiv:1904.04493 [hep-ph]} \BibitemShut {NoStop}%
\bibitem [{\citenamefont {Tang}\ and\ \citenamefont
  {Wu}(2020{\natexlab{c}})}]{Tang:2019olx}%
  \BibitemOpen
  \bibfield  {author} {\bibinfo {author} {\bibfnamefont {Y.}~\bibnamefont
  {Tang}}\ and\ \bibinfo {author} {\bibfnamefont {Y.-L.}\ \bibnamefont {Wu}},\
  }\bibfield  {title} {\bibinfo {title} {{Conformal $\alpha$-attractor
  inflation with Weyl gauge field}},\ }\href
  {https://doi.org/10.1088/1475-7516/2020/03/067} {\bibfield  {journal}
  {\bibinfo  {journal} {JCAP}\ }\textbf {\bibinfo {volume} {03}},\ \bibinfo
  {pages} {067}},\ \Eprint {https://arxiv.org/abs/1912.07610} {arXiv:1912.07610
  [hep-ph]} \BibitemShut {NoStop}%
\bibitem [{\citenamefont {Ghilencea}(2021)}]{Ghilencea:2020rxc}%
  \BibitemOpen
  \bibfield  {author} {\bibinfo {author} {\bibfnamefont {D.~M.}\ \bibnamefont
  {Ghilencea}},\ }\bibfield  {title} {\bibinfo {title} {{Gauging scale symmetry
  and inflation: Weyl versus Palatini gravity}},\ }\href
  {https://doi.org/10.1140/epjc/s10052-021-09226-1} {\bibfield  {journal}
  {\bibinfo  {journal} {Eur. Phys. J. C}\ }\textbf {\bibinfo {volume} {81}},\
  \bibinfo {pages} {510} (\bibinfo {year} {2021})},\ \Eprint
  {https://arxiv.org/abs/2007.14733} {arXiv:2007.14733 [hep-th]} \BibitemShut
  {NoStop}%
\bibitem [{\citenamefont {Ghilencea}\ and\ \citenamefont
  {Harko}(2021)}]{Ghilencea:2021jjl}%
  \BibitemOpen
  \bibfield  {author} {\bibinfo {author} {\bibfnamefont {D.~M.}\ \bibnamefont
  {Ghilencea}}\ and\ \bibinfo {author} {\bibfnamefont {T.}~\bibnamefont
  {Harko}},\ }\bibfield  {title} {\bibinfo {title} {{Cosmological evolution in
  Weyl conformal geometry}},\ }\href@noop {} {\  (\bibinfo {year} {2021})},\
  \Eprint {https://arxiv.org/abs/2110.07056} {arXiv:2110.07056 [gr-qc]}
  \BibitemShut {NoStop}%
\bibitem [{\citenamefont {Wang}\ \emph {et~al.}(2022)\citenamefont {Wang},
  \citenamefont {Tang},\ and\ \citenamefont {Wu}}]{Wang:2022ojc}%
  \BibitemOpen
  \bibfield  {author} {\bibinfo {author} {\bibfnamefont {Q.-Y.}\ \bibnamefont
  {Wang}}, \bibinfo {author} {\bibfnamefont {Y.}~\bibnamefont {Tang}},\ and\
  \bibinfo {author} {\bibfnamefont {Y.-L.}\ \bibnamefont {Wu}},\ }\bibfield
  {title} {\bibinfo {title} {{Dark matter production in Weyl $R^2$
  inflation}},\ }\href {https://doi.org/10.1103/PhysRevD.106.023502} {\bibfield
   {journal} {\bibinfo  {journal} {Phys. Rev. D}\ }\textbf {\bibinfo {volume}
  {106}},\ \bibinfo {pages} {023502} (\bibinfo {year} {2022})},\ \Eprint
  {https://arxiv.org/abs/2203.15452} {arXiv:2203.15452 [hep-ph]} \BibitemShut
  {NoStop}%
\bibitem [{\citenamefont {Wang}\ \emph {et~al.}(2023)\citenamefont {Wang},
  \citenamefont {Tang},\ and\ \citenamefont {Wu}}]{Wang:2023hsb}%
  \BibitemOpen
  \bibfield  {author} {\bibinfo {author} {\bibfnamefont {Q.-Y.}\ \bibnamefont
  {Wang}}, \bibinfo {author} {\bibfnamefont {Y.}~\bibnamefont {Tang}},\ and\
  \bibinfo {author} {\bibfnamefont {Y.-L.}\ \bibnamefont {Wu}},\ }\bibfield
  {title} {\bibinfo {title} {{Inflation in Weyl scaling invariant gravity with
  $R^3$ extensions}},\ }\href {https://doi.org/10.1103/PhysRevD.107.083511}
  {\bibfield  {journal} {\bibinfo  {journal} {Phys. Rev. D}\ }\textbf {\bibinfo
  {volume} {107}},\ \bibinfo {pages} {083511} (\bibinfo {year} {2023})},\
  \Eprint {https://arxiv.org/abs/2301.03744} {arXiv:2301.03744 [astro-ph.CO]}
  \BibitemShut {NoStop}%
\bibitem [{\citenamefont {Hu}\ \emph {et~al.}(2024)\citenamefont {Hu},
  \citenamefont {Wang}, \citenamefont {Ma},\ and\ \citenamefont
  {Tang}}]{Hu:2023yjn}%
  \BibitemOpen
  \bibfield  {author} {\bibinfo {author} {\bibfnamefont {W.-Y.}\ \bibnamefont
  {Hu}}, \bibinfo {author} {\bibfnamefont {Q.-Y.}\ \bibnamefont {Wang}},
  \bibinfo {author} {\bibfnamefont {Y.-Q.}\ \bibnamefont {Ma}},\ and\ \bibinfo
  {author} {\bibfnamefont {Y.}~\bibnamefont {Tang}},\ }\bibfield  {title}
  {\bibinfo {title} {{Gravitational waves from preheating in inflation with
  Weyl symmetry}},\ }\href {https://doi.org/10.1103/PhysRevD.109.083542}
  {\bibfield  {journal} {\bibinfo  {journal} {Phys. Rev. D}\ }\textbf {\bibinfo
  {volume} {109}},\ \bibinfo {pages} {083542} (\bibinfo {year} {2024})},\
  \Eprint {https://arxiv.org/abs/2311.00239} {arXiv:2311.00239 [astro-ph.CO]}
  \BibitemShut {NoStop}%
\bibitem [{\citenamefont {Burikham}\ \emph {et~al.}(2023)\citenamefont
  {Burikham}, \citenamefont {Harko}, \citenamefont {Pimsamarn},\ and\
  \citenamefont {Shahidi}}]{Burikham:2023bil}%
  \BibitemOpen
  \bibfield  {author} {\bibinfo {author} {\bibfnamefont {P.}~\bibnamefont
  {Burikham}}, \bibinfo {author} {\bibfnamefont {T.}~\bibnamefont {Harko}},
  \bibinfo {author} {\bibfnamefont {K.}~\bibnamefont {Pimsamarn}},\ and\
  \bibinfo {author} {\bibfnamefont {S.}~\bibnamefont {Shahidi}},\ }\bibfield
  {title} {\bibinfo {title} {{Dark matter as a Weyl geometric effect}},\ }\href
  {https://doi.org/10.1103/PhysRevD.107.064008} {\bibfield  {journal} {\bibinfo
   {journal} {Phys. Rev. D}\ }\textbf {\bibinfo {volume} {107}},\ \bibinfo
  {pages} {064008} (\bibinfo {year} {2023})},\ \Eprint
  {https://arxiv.org/abs/2302.08289} {arXiv:2302.08289 [gr-qc]} \BibitemShut
  {NoStop}%
\bibitem [{\citenamefont {Bhagat}\ \emph {et~al.}(2023)\citenamefont {Bhagat},
  \citenamefont {Narawade},\ and\ \citenamefont {Mishra}}]{Bhagat:2023ych}%
  \BibitemOpen
  \bibfield  {author} {\bibinfo {author} {\bibfnamefont {R.}~\bibnamefont
  {Bhagat}}, \bibinfo {author} {\bibfnamefont {S.~A.}\ \bibnamefont
  {Narawade}},\ and\ \bibinfo {author} {\bibfnamefont {B.}~\bibnamefont
  {Mishra}},\ }\bibfield  {title} {\bibinfo {title} {{Weyl type f(Q,T) gravity
  observational constrained cosmological models}},\ }\href
  {https://doi.org/10.1016/j.dark.2023.101250} {\bibfield  {journal} {\bibinfo
  {journal} {Phys. Dark Univ.}\ }\textbf {\bibinfo {volume} {41}},\ \bibinfo
  {pages} {101250} (\bibinfo {year} {2023})},\ \Eprint
  {https://arxiv.org/abs/2305.01659} {arXiv:2305.01659 [gr-qc]} \BibitemShut
  {NoStop}%
\bibitem [{\citenamefont {Harko}\ and\ \citenamefont
  {Shahidi}(2024)}]{Harko:2024fnt}%
  \BibitemOpen
  \bibfield  {author} {\bibinfo {author} {\bibfnamefont {T.}~\bibnamefont
  {Harko}}\ and\ \bibinfo {author} {\bibfnamefont {S.}~\bibnamefont
  {Shahidi}},\ }\bibfield  {title} {\bibinfo {title} {{Cosmological
  implications of the Weyl geometric gravity theory}},\ }\href
  {https://doi.org/10.1140/epjc/s10052-024-12861-z} {\bibfield  {journal}
  {\bibinfo  {journal} {Eur. Phys. J. C}\ }\textbf {\bibinfo {volume} {84}},\
  \bibinfo {pages} {509} (\bibinfo {year} {2024})},\ \Eprint
  {https://arxiv.org/abs/2405.04129} {arXiv:2405.04129 [gr-qc]} \BibitemShut
  {NoStop}%
\bibitem [{\citenamefont {Gialamas}\ and\ \citenamefont
  {Tamvakis}(2025)}]{Gialamas:2024iyu}%
  \BibitemOpen
  \bibfield  {author} {\bibinfo {author} {\bibfnamefont {I.~D.}\ \bibnamefont
  {Gialamas}}\ and\ \bibinfo {author} {\bibfnamefont {K.}~\bibnamefont
  {Tamvakis}},\ }\bibfield  {title} {\bibinfo {title} {{Inflation in
  Weyl-invariant Einstein-Cartan gravity}},\ }\href
  {https://doi.org/10.1103/PhysRevD.111.044007} {\bibfield  {journal} {\bibinfo
   {journal} {Phys. Rev. D}\ }\textbf {\bibinfo {volume} {111}},\ \bibinfo
  {pages} {044007} (\bibinfo {year} {2025})},\ \Eprint
  {https://arxiv.org/abs/2410.16364} {arXiv:2410.16364 [gr-qc]} \BibitemShut
  {NoStop}%
\bibitem [{\citenamefont {Lee}(2025)}]{Lee:2024rjw}%
  \BibitemOpen
  \bibfield  {author} {\bibinfo {author} {\bibfnamefont {H.~M.}\ \bibnamefont
  {Lee}},\ }\bibfield  {title} {\bibinfo {title} {{Pole Inflation from Broken
  Noncompact Isometry in Weyl Gravity}},\ }\href
  {https://doi.org/10.1103/skhx-yc43} {\bibfield  {journal} {\bibinfo
  {journal} {Phys. Rev. Lett.}\ }\textbf {\bibinfo {volume} {134}},\ \bibinfo
  {pages} {211001} (\bibinfo {year} {2025})},\ \Eprint
  {https://arxiv.org/abs/2411.16944} {arXiv:2411.16944 [hep-ph]} \BibitemShut
  {NoStop}%
\bibitem [{\citenamefont {Konoplya}\ \emph {et~al.}(2025)\citenamefont
  {Konoplya}, \citenamefont {Khrabustovskyi}, \citenamefont
  {K{\v{r}}{\'\i}{\v{z}}},\ and\ \citenamefont {Zhidenko}}]{Konoplya:2025mvj}%
  \BibitemOpen
  \bibfield  {author} {\bibinfo {author} {\bibfnamefont {R.~A.}\ \bibnamefont
  {Konoplya}}, \bibinfo {author} {\bibfnamefont {A.}~\bibnamefont
  {Khrabustovskyi}}, \bibinfo {author} {\bibfnamefont {J.}~\bibnamefont
  {K{\v{r}}{\'\i}{\v{z}}}},\ and\ \bibinfo {author} {\bibfnamefont
  {A.}~\bibnamefont {Zhidenko}},\ }\bibfield  {title} {\bibinfo {title}
  {{Quasinormal ringing and shadows of black holes and wormholes in dark
  matter-inspired Weyl gravity}},\ }\href
  {https://doi.org/10.1088/1475-7516/2025/04/062} {\bibfield  {journal}
  {\bibinfo  {journal} {JCAP}\ }\textbf {\bibinfo {volume} {04}},\ \bibinfo
  {pages} {062}},\ \Eprint {https://arxiv.org/abs/2501.16134} {arXiv:2501.16134
  [gr-qc]} \BibitemShut {NoStop}%
\bibitem [{\citenamefont {Lalak}\ and\ \citenamefont
  {Michalak}(2026)}]{Lalak:2025mil}%
  \BibitemOpen
  \bibfield  {author} {\bibinfo {author} {\bibfnamefont {Z.}~\bibnamefont
  {Lalak}}\ and\ \bibinfo {author} {\bibfnamefont {P.}~\bibnamefont
  {Michalak}},\ }\bibfield  {title} {\bibinfo {title} {{Inflation in the scale
  symmetric Standard Model and Weyl geometry}},\ }\href
  {https://doi.org/10.1007/JHEP02(2026)041} {\bibfield  {journal} {\bibinfo
  {journal} {JHEP}\ }\textbf {\bibinfo {volume} {02}},\ \bibinfo {pages}
  {041}},\ \Eprint {https://arxiv.org/abs/2506.14617} {arXiv:2506.14617
  [hep-ph]} \BibitemShut {NoStop}%
\bibitem [{\citenamefont {Katsoulas}\ and\ \citenamefont
  {Tamvakis}(2026)}]{Katsoulas:2025srh}%
  \BibitemOpen
  \bibfield  {author} {\bibinfo {author} {\bibfnamefont {T.}~\bibnamefont
  {Katsoulas}}\ and\ \bibinfo {author} {\bibfnamefont {K.}~\bibnamefont
  {Tamvakis}},\ }\bibfield  {title} {\bibinfo {title} {{Inflationary assessment
  of F(R,R{\textasciitilde}) Einstein-Cartan models}},\ }\href
  {https://doi.org/10.1103/zhh9-t74c} {\bibfield  {journal} {\bibinfo
  {journal} {Phys. Rev. D}\ }\textbf {\bibinfo {volume} {113}},\ \bibinfo
  {pages} {124010} (\bibinfo {year} {2026})},\ \Eprint
  {https://arxiv.org/abs/2512.02847} {arXiv:2512.02847 [gr-qc]} \BibitemShut
  {NoStop}%
\bibitem [{\citenamefont {Wu}(2016)}]{Wu:2015wwa}%
  \BibitemOpen
  \bibfield  {author} {\bibinfo {author} {\bibfnamefont {Y.-L.}\ \bibnamefont
  {Wu}},\ }\bibfield  {title} {\bibinfo {title} {{Quantum field theory of
  gravity with spin and scaling gauge invariance and spacetime dynamics with
  quantum inflation}},\ }\href {https://doi.org/10.1103/PhysRevD.93.024012}
  {\bibfield  {journal} {\bibinfo  {journal} {Phys. Rev. D}\ }\textbf {\bibinfo
  {volume} {93}},\ \bibinfo {pages} {024012} (\bibinfo {year} {2016})},\
  \Eprint {https://arxiv.org/abs/1506.01807} {arXiv:1506.01807 [hep-th]}
  \BibitemShut {NoStop}%
\bibitem [{\citenamefont {Ghilencea}(2025)}]{Ghilencea:2024usf}%
  \BibitemOpen
  \bibfield  {author} {\bibinfo {author} {\bibfnamefont {D.~M.}\ \bibnamefont
  {Ghilencea}},\ }\bibfield  {title} {\bibinfo {title} {{Quantum gravity from
  Weyl conformal geometry}},\ }\href
  {https://doi.org/10.1140/epjc/s10052-025-14489-z} {\bibfield  {journal}
  {\bibinfo  {journal} {Eur. Phys. J. C}\ }\textbf {\bibinfo {volume} {85}},\
  \bibinfo {pages} {815} (\bibinfo {year} {2025})},\ \Eprint
  {https://arxiv.org/abs/2408.07160} {arXiv:2408.07160 [hep-th]} \BibitemShut
  {NoStop}%
\bibitem [{\citenamefont {Huang}(2014)}]{Huang:2013hsb}%
  \BibitemOpen
  \bibfield  {author} {\bibinfo {author} {\bibfnamefont {Q.-G.}\ \bibnamefont
  {Huang}},\ }\bibfield  {title} {\bibinfo {title} {{A polynomial f(R)
  inflation model}},\ }\href {https://doi.org/10.1088/1475-7516/2014/02/035}
  {\bibfield  {journal} {\bibinfo  {journal} {JCAP}\ }\textbf {\bibinfo
  {volume} {02}},\ \bibinfo {pages} {035}},\ \Eprint
  {https://arxiv.org/abs/1309.3514} {arXiv:1309.3514 [hep-th]} \BibitemShut
  {NoStop}%
\bibitem [{\citenamefont {Asaka}\ \emph {et~al.}(2016)\citenamefont {Asaka},
  \citenamefont {Iso}, \citenamefont {Kawai}, \citenamefont {Kohri},
  \citenamefont {Noumi},\ and\ \citenamefont {Terada}}]{Asaka:2015vza}%
  \BibitemOpen
  \bibfield  {author} {\bibinfo {author} {\bibfnamefont {T.}~\bibnamefont
  {Asaka}}, \bibinfo {author} {\bibfnamefont {S.}~\bibnamefont {Iso}}, \bibinfo
  {author} {\bibfnamefont {H.}~\bibnamefont {Kawai}}, \bibinfo {author}
  {\bibfnamefont {K.}~\bibnamefont {Kohri}}, \bibinfo {author} {\bibfnamefont
  {T.}~\bibnamefont {Noumi}},\ and\ \bibinfo {author} {\bibfnamefont
  {T.}~\bibnamefont {Terada}},\ }\bibfield  {title} {\bibinfo {title}
  {{Reinterpretation of the Starobinsky model}},\ }\href
  {https://doi.org/10.1093/ptep/ptw161} {\bibfield  {journal} {\bibinfo
  {journal} {PTEP}\ }\textbf {\bibinfo {volume} {2016}},\ \bibinfo {pages}
  {123E01} (\bibinfo {year} {2016})},\ \Eprint
  {https://arxiv.org/abs/1507.04344} {arXiv:1507.04344 [hep-th]} \BibitemShut
  {NoStop}%
\bibitem [{\citenamefont {Cheong}\ \emph {et~al.}(2020)\citenamefont {Cheong},
  \citenamefont {Lee},\ and\ \citenamefont {Park}}]{Cheong:2020rao}%
  \BibitemOpen
  \bibfield  {author} {\bibinfo {author} {\bibfnamefont {D.~Y.}\ \bibnamefont
  {Cheong}}, \bibinfo {author} {\bibfnamefont {H.~M.}\ \bibnamefont {Lee}},\
  and\ \bibinfo {author} {\bibfnamefont {S.~C.}\ \bibnamefont {Park}},\
  }\bibfield  {title} {\bibinfo {title} {{Beyond the Starobinsky model for
  inflation}},\ }\href {https://doi.org/10.1016/j.physletb.2020.135453}
  {\bibfield  {journal} {\bibinfo  {journal} {Phys. Lett. B}\ }\textbf
  {\bibinfo {volume} {805}},\ \bibinfo {pages} {135453} (\bibinfo {year}
  {2020})},\ \Eprint {https://arxiv.org/abs/2002.07981} {arXiv:2002.07981
  [hep-ph]} \BibitemShut {NoStop}%
\bibitem [{\citenamefont {Rodrigues-da Silva}\ \emph
  {et~al.}(2022)\citenamefont {Rodrigues-da Silva}, \citenamefont
  {Bezerra-Sobrinho},\ and\ \citenamefont
  {Medeiros}}]{Rodrigues-da-Silva:2021jab}%
  \BibitemOpen
  \bibfield  {author} {\bibinfo {author} {\bibfnamefont {G.}~\bibnamefont
  {Rodrigues-da Silva}}, \bibinfo {author} {\bibfnamefont {J.}~\bibnamefont
  {Bezerra-Sobrinho}},\ and\ \bibinfo {author} {\bibfnamefont {L.~G.}\
  \bibnamefont {Medeiros}},\ }\bibfield  {title} {\bibinfo {title}
  {{Higher-order extension of Starobinsky inflation: Initial conditions,
  slow-roll regime, and reheating phase}},\ }\href
  {https://doi.org/10.1103/PhysRevD.105.063504} {\bibfield  {journal} {\bibinfo
   {journal} {Phys. Rev. D}\ }\textbf {\bibinfo {volume} {105}},\ \bibinfo
  {pages} {063504} (\bibinfo {year} {2022})},\ \Eprint
  {https://arxiv.org/abs/2110.15502} {arXiv:2110.15502 [astro-ph.CO]}
  \BibitemShut {NoStop}%
\bibitem [{\citenamefont {Ivanov}\ \emph {et~al.}(2022)\citenamefont {Ivanov},
  \citenamefont {Ketov}, \citenamefont {Pozdeeva},\ and\ \citenamefont
  {Vernov}}]{Ivanov:2021chn}%
  \BibitemOpen
  \bibfield  {author} {\bibinfo {author} {\bibfnamefont {V.~R.}\ \bibnamefont
  {Ivanov}}, \bibinfo {author} {\bibfnamefont {S.~V.}\ \bibnamefont {Ketov}},
  \bibinfo {author} {\bibfnamefont {E.~O.}\ \bibnamefont {Pozdeeva}},\ and\
  \bibinfo {author} {\bibfnamefont {S.~Y.}\ \bibnamefont {Vernov}},\ }\bibfield
   {title} {\bibinfo {title} {{Analytic extensions of Starobinsky model of
  inflation}},\ }\href {https://doi.org/10.1088/1475-7516/2022/03/058}
  {\bibfield  {journal} {\bibinfo  {journal} {JCAP}\ }\textbf {\bibinfo
  {volume} {03}}\bibfield  {number} {\bibinfo  {number} { (03)},\ \bibinfo
  {pages} {058}},\ }\Eprint {https://arxiv.org/abs/2111.09058}
  {arXiv:2111.09058 [gr-qc]} \BibitemShut {NoStop}%
\bibitem [{\citenamefont {Shtanov}\ \emph {et~al.}(2023)\citenamefont
  {Shtanov}, \citenamefont {Sahni},\ and\ \citenamefont
  {Mishra}}]{Shtanov:2022pdx}%
  \BibitemOpen
  \bibfield  {author} {\bibinfo {author} {\bibfnamefont {Y.}~\bibnamefont
  {Shtanov}}, \bibinfo {author} {\bibfnamefont {V.}~\bibnamefont {Sahni}},\
  and\ \bibinfo {author} {\bibfnamefont {S.~S.}\ \bibnamefont {Mishra}},\
  }\bibfield  {title} {\bibinfo {title} {{Tabletop potentials for inflation
  from f(R) gravity}},\ }\href {https://doi.org/10.1088/1475-7516/2023/03/023}
  {\bibfield  {journal} {\bibinfo  {journal} {JCAP}\ }\textbf {\bibinfo
  {volume} {03}},\ \bibinfo {pages} {023}},\ \Eprint
  {https://arxiv.org/abs/2210.01828} {arXiv:2210.01828 [gr-qc]} \BibitemShut
  {NoStop}%
\bibitem [{\citenamefont {Modak}\ \emph {et~al.}(2023)\citenamefont {Modak},
  \citenamefont {R\"over}, \citenamefont {Sch\"afer}, \citenamefont
  {Schosser},\ and\ \citenamefont {Plehn}}]{Modak:2022gol}%
  \BibitemOpen
  \bibfield  {author} {\bibinfo {author} {\bibfnamefont {T.}~\bibnamefont
  {Modak}}, \bibinfo {author} {\bibfnamefont {L.}~\bibnamefont {R\"over}},
  \bibinfo {author} {\bibfnamefont {B.~M.}\ \bibnamefont {Sch\"afer}}, \bibinfo
  {author} {\bibfnamefont {B.}~\bibnamefont {Schosser}},\ and\ \bibinfo
  {author} {\bibfnamefont {T.}~\bibnamefont {Plehn}},\ }\bibfield  {title}
  {\bibinfo {title} {{Cornering extended Starobinsky inflation with CMB and
  SKA}},\ }\href {https://doi.org/10.21468/SciPostPhys.15.2.047} {\bibfield
  {journal} {\bibinfo  {journal} {SciPost Phys.}\ }\textbf {\bibinfo {volume}
  {15}},\ \bibinfo {pages} {047} (\bibinfo {year} {2023})},\ \Eprint
  {https://arxiv.org/abs/2210.05698} {arXiv:2210.05698 [astro-ph.CO]}
  \BibitemShut {NoStop}%
\bibitem [{\citenamefont {Gialamas}\ \emph {et~al.}(2025)\citenamefont
  {Gialamas}, \citenamefont {Katsoulas},\ and\ \citenamefont
  {Tamvakis}}]{Gialamas:2025ofz}%
  \BibitemOpen
  \bibfield  {author} {\bibinfo {author} {\bibfnamefont {I.~D.}\ \bibnamefont
  {Gialamas}}, \bibinfo {author} {\bibfnamefont {T.}~\bibnamefont
  {Katsoulas}},\ and\ \bibinfo {author} {\bibfnamefont {K.}~\bibnamefont
  {Tamvakis}},\ }\bibfield  {title} {\bibinfo {title} {{Keeping the relation
  between the Starobinsky model and no-scale supergravity ACTive}},\ }\href
  {https://doi.org/10.1088/1475-7516/2025/09/060} {\bibfield  {journal}
  {\bibinfo  {journal} {JCAP}\ }\textbf {\bibinfo {volume} {09}},\ \bibinfo
  {pages} {060}},\ \Eprint {https://arxiv.org/abs/2505.03608} {arXiv:2505.03608
  [gr-qc]} \BibitemShut {NoStop}%
\bibitem [{\citenamefont {Odintsov}\ and\ \citenamefont
  {Oikonomou}(2025)}]{Odintsov:2025eiv}%
  \BibitemOpen
  \bibfield  {author} {\bibinfo {author} {\bibfnamefont {S.~D.}\ \bibnamefont
  {Odintsov}}\ and\ \bibinfo {author} {\bibfnamefont {V.~K.}\ \bibnamefont
  {Oikonomou}},\ }\bibfield  {title} {\bibinfo {title} {{Power-law F(R) gravity
  as deformations to Starobinsky inflation in view of ACT}},\ }\href
  {https://doi.org/10.1016/j.physletb.2025.139907} {\bibfield  {journal}
  {\bibinfo  {journal} {Phys. Lett. B}\ }\textbf {\bibinfo {volume} {870}},\
  \bibinfo {pages} {139907} (\bibinfo {year} {2025})},\ \Eprint
  {https://arxiv.org/abs/2509.06251} {arXiv:2509.06251 [gr-qc]} \BibitemShut
  {NoStop}%
\bibitem [{\citenamefont {Qiu}\ \emph {et~al.}(2026)\citenamefont {Qiu},
  \citenamefont {Pang},\ and\ \citenamefont {Huang}}]{Qiu:2025iqm}%
  \BibitemOpen
  \bibfield  {author} {\bibinfo {author} {\bibfnamefont {Z.}~\bibnamefont
  {Qiu}}, \bibinfo {author} {\bibfnamefont {Y.}~\bibnamefont {Pang}},\ and\
  \bibinfo {author} {\bibfnamefont {Q.}~\bibnamefont {Huang}},\ }\bibfield
  {title} {\bibinfo {title} {{The implications of inflation for the last
  ACT}},\ }\href {https://doi.org/10.1007/s11433-025-2934-8} {\bibfield
  {journal} {\bibinfo  {journal} {Sci. China Phys. Mech. Astron.}\ }\textbf
  {\bibinfo {volume} {69}},\ \bibinfo {pages} {260413} (\bibinfo {year}
  {2026})},\ \Eprint {https://arxiv.org/abs/2510.18320} {arXiv:2510.18320
  [astro-ph.CO]} \BibitemShut {NoStop}%
\bibitem [{\citenamefont {Bezerra-Sobrinho}\ and\ \citenamefont
  {Medeiros}(2026)}]{Bezerra-Sobrinho:2025gfg}%
  \BibitemOpen
  \bibfield  {author} {\bibinfo {author} {\bibfnamefont {J.}~\bibnamefont
  {Bezerra-Sobrinho}}\ and\ \bibinfo {author} {\bibfnamefont {L.~G.}\
  \bibnamefont {Medeiros}},\ }\bibfield  {title} {\bibinfo {title}
  {{Starobinsky inflation and the latest CMB data: a subtle tension?}},\ }\href
  {https://doi.org/10.1140/epjc/s10052-026-15686-0} {\bibfield  {journal}
  {\bibinfo  {journal} {Eur. Phys. J. C}\ }\textbf {\bibinfo {volume} {86}},\
  \bibinfo {pages} {416} (\bibinfo {year} {2026})},\ \Eprint
  {https://arxiv.org/abs/2511.06640} {arXiv:2511.06640 [astro-ph.CO]}
  \BibitemShut {NoStop}%
\bibitem [{\citenamefont {Graham}\ \emph {et~al.}(2016)\citenamefont {Graham},
  \citenamefont {Mardon},\ and\ \citenamefont {Rajendran}}]{Graham:2015rva}%
  \BibitemOpen
  \bibfield  {author} {\bibinfo {author} {\bibfnamefont {P.~W.}\ \bibnamefont
  {Graham}}, \bibinfo {author} {\bibfnamefont {J.}~\bibnamefont {Mardon}},\
  and\ \bibinfo {author} {\bibfnamefont {S.}~\bibnamefont {Rajendran}},\
  }\bibfield  {title} {\bibinfo {title} {{Vector Dark Matter from Inflationary
  Fluctuations}},\ }\href {https://doi.org/10.1103/PhysRevD.93.103520}
  {\bibfield  {journal} {\bibinfo  {journal} {Phys. Rev. D}\ }\textbf {\bibinfo
  {volume} {93}},\ \bibinfo {pages} {103520} (\bibinfo {year} {2016})},\
  \Eprint {https://arxiv.org/abs/1504.02102} {arXiv:1504.02102 [hep-ph]}
  \BibitemShut {NoStop}%
\bibitem [{\citenamefont {Ema}\ \emph {et~al.}(2019)\citenamefont {Ema},
  \citenamefont {Nakayama},\ and\ \citenamefont {Tang}}]{Ema:2019yrd}%
  \BibitemOpen
  \bibfield  {author} {\bibinfo {author} {\bibfnamefont {Y.}~\bibnamefont
  {Ema}}, \bibinfo {author} {\bibfnamefont {K.}~\bibnamefont {Nakayama}},\ and\
  \bibinfo {author} {\bibfnamefont {Y.}~\bibnamefont {Tang}},\ }\bibfield
  {title} {\bibinfo {title} {{Production of purely gravitational dark matter:
  the case of fermion and vector boson}},\ }\href
  {https://doi.org/10.1007/JHEP07(2019)060} {\bibfield  {journal} {\bibinfo
  {journal} {JHEP}\ }\textbf {\bibinfo {volume} {07}},\ \bibinfo {pages}
  {060}},\ \Eprint {https://arxiv.org/abs/1903.10973} {arXiv:1903.10973
  [hep-ph]} \BibitemShut {NoStop}%
\bibitem [{\citenamefont {Ahmed}\ \emph {et~al.}(2020)\citenamefont {Ahmed},
  \citenamefont {Grzadkowski},\ and\ \citenamefont {Socha}}]{Ahmed:2020fhc}%
  \BibitemOpen
  \bibfield  {author} {\bibinfo {author} {\bibfnamefont {A.}~\bibnamefont
  {Ahmed}}, \bibinfo {author} {\bibfnamefont {B.}~\bibnamefont {Grzadkowski}},\
  and\ \bibinfo {author} {\bibfnamefont {A.}~\bibnamefont {Socha}},\ }\bibfield
   {title} {\bibinfo {title} {{Gravitational production of vector dark
  matter}},\ }\href {https://doi.org/10.1007/JHEP08(2020)059} {\bibfield
  {journal} {\bibinfo  {journal} {JHEP}\ }\textbf {\bibinfo {volume} {08}},\
  \bibinfo {pages} {059}},\ \Eprint {https://arxiv.org/abs/2005.01766}
  {arXiv:2005.01766 [hep-ph]} \BibitemShut {NoStop}%
\end{thebibliography}%

\end{document}